\documentclass[a4paper,11pt]{article}
\pdfoutput=1
\usepackage[utf8]{inputenc}
\usepackage[T1]{fontenc}
\usepackage{jheppub}
\usepackage{natbib}

\usepackage[usenames,dvipsnames,svgnames,table,x11names]{xcolor}
\usepackage{lmodern,amsmath,multirow,cancel}
\usepackage{booktabs}
\usepackage{xstring}
\usepackage{ifthen}

\usepackage{ascmac,braket,bm,mathrsfs,amsthm,amsfonts}
\usepackage{hyperref}
\usepackage{graphicx}
\usepackage{subcaption} 
\usepackage{comment}
\usepackage[normalem]{ulem}

\usepackage{titlesec}
\newcommand*{\justifyheading}{\raggedright}
\titleformat{\chapter}[display]
  {\normalfont\huge\bfseries\justifyheading}{\chaptertitlename\ \thechapter}
  {20pt}{\Huge}
\titleformat{\section}
  {\normalfont\Large\bfseries\justifyheading}{\thesection}{1em}{}
\titleformat{\subsection}
  {\normalfont\large\bfseries\justifyheading}{\thesubsection}{1em}{}
\titleformat{\subsubsection}
  {\normalfont\bfseries\justifyheading}{\thesubsubsection}{1em}{}
\usepackage[english]{babel}
\usepackage[dvipsnames]{xcolor}
\numberwithin{equation}{section}

\usepackage{slashed}

\DeclareMathOperator{\Tr}{\mathrm{Tr}}

\renewcommand\Im{\mathop{\mathrm{Im}}}

\newcommand\unit[1]{\,\mathrm{#1}}

\usepackage{environ}

\newcommand{\lsim}{ \mathop{}_{\textstyle \sim}^{\textstyle <} }

\NewEnviron{eqsp}{%
\begin{equation}\begin{split}
  \BODY
\end{split}\end{equation}
}
\renewcommand{\proofname}{$\because$}


\usepackage{adjustbox}
\usepackage{tabularx}
\newcolumntype{Y}{>{\centering\arraybackslash}X} 
\usepackage{booktabs} 
\usepackage{colortbl} 
\def\beq#1\eeq{\begin{align}#1\end{align}}

\RequirePackage{xspace}
\def\Bbar    {\kern 0.18em\overline{\kern -0.18em B}{}\xspace}

\newcommand{\bhline}[1]{\noalign{\hrule height #1}}

\preprint{CHIBA-EP-284, IPMU26-0036}

\title{Radiative \boldmath{$\theta$}  in Left-Right Universal Seesaw}

\author[a]{Tatsuya Banno,}
\author[a,b,c]{Junji Hisano,}
\author[d,b]{Teppei Kitahara,}
\author[e]{and\\ Naohiro Osamura}

\affiliation[a]{
  Department of Physics, Nagoya University, Furo-cho Chikusa-ku, Nagoya 464-8602, Japan
}
\affiliation[b]{
  Kobayashi-Maskawa Institute for the Origin of Particles and the
  Universe, Nagoya University,
  Furo-cho Chikusa-ku, Nagoya 464-8602, Japan
}
\affiliation[c]{
  Kavli IPMU (WPI), UTIAS, The University of Tokyo, Kashiwa 277-8584, Japan
}

\affiliation[d]{
Department of Physics, Graduate School of Science,
Chiba University, Chiba 263-8522, Japan}

\affiliation[e]{
ICRR, The University of Tokyo, Kashiwa, Chiba 277-8582, Japan
}

\emailAdd{banno.tatsuya.p8@s.mail.nagoya-u.ac.jp}
\emailAdd{hisano@eken.phys.nagoya-u.ac.jp}
\emailAdd{kitahara@chiba-u.jp}
\emailAdd{osamura@icrr.u-tokyo.ac.jp}

\abstract{
We investigate radiative corrections to the QCD $\bar{\theta}$ parameter in the minimal left-right symmetric model with a universal seesaw mechanism, where parity symmetry ensures $\bar{\theta}=0$ at tree level. Assuming Hermitian vector-like quark mass matrices, we identify the $CP$-odd rephasing invariants contributing to $\bar{\theta}$ and evaluate the leading two-loop corrections to the light-quark mass matrices induced by soft parity breaking. 
We find that the dominant contributions do not decouple
and depend sensitively on the mass splittings as well as flavor mixings of the vector-like quarks. 
For hierarchical vector-like quark masses, the resulting $\bar{\theta}$ typically exceeds the current experimental bound from the neutron electric dipole moment. 
Satisfying this bound requires either 
nearly degenerate vector-like quark masses or
strongly suppressed additional flavor mixing angles.
In the former case,
the radiative corrections are suppressed by a GIM-like mechanism. 
For generic mixing parameters, the dimensionless relative mass splittings are constrained to be below approximately $10^{-2}$ and $10^{-4}$ in the up- and down-type vector-like quark sectors, respectively. These results severely restrict the inverted vector-like quark mass hierarchy motivated by the universal seesaw mechanism.}


\begin{document}

\sloppy 

\makeatletter\renewcommand{\@fpheader}{\ }\makeatother

\maketitle

\renewcommand{\thefootnote}{\#\arabic{footnote}}
\setcounter{footnote}{0}


\section{Introduction}
\label{sec:intro}

It is well established that the quark-QCD sector of the Standard Model (SM) contains two independent sources of physical $CP$ violation, after removing all unphysical phases by quark-field redefinitions.
One is the $CP$-violating phase of the Cabibbo-Kobayashi-Maskawa (CKM) matrix \cite{Kobayashi:1973fv},
which has been precisely measured to be $\delta_\mathrm{CKM}=1.154\pm0.025$ through  flavor physics~\cite{ParticleDataGroup:2026mpi}.
The other is the QCD $\theta$ parameter, defined by the rephasing-invariant combination
\begin{align}
  \bar{\theta} = \theta_G - \theta_q\,,
\label{eq1}
\end{align}
where $\theta_G$ is the coefficient of the bare QCD $\theta$ term and $\theta_q$ is the phase of the determinant of the quark mass matrix~\cite{Adler:1969gk,Bell:1969ts,Adler:1969er,tHooft:1972tcz,Fujikawa:1979ay}.
 A nonzero $\bar{\theta}$ induces $CP$ violation in the strong sector, 
 conventionally referred to as strong $CP$ violation.
 The CKM phase induces a nonzero $\bar{\theta}$ through higher-order radiative corrections in the SM, although its estimated value is extremely small, $\bar{\theta}\simeq 10^{-19}$~\cite{Khriplovich:1985jr,Pospelov:1994uf}.

The neutron electric dipole moment (nEDM) provides 
one of the most sensitive probes of $\bar{\theta}$.
Combining the current nEDM upper bound, $|d_n| < 1.8\times10^{-26}\,e\unit{cm}$~\cite{Abel:2020pzs}, 
with the lattice QCD result $d_n = -(1.48\pm0.34)\times10^{-16} \,\bar{\theta}\, e\unit{cm}$~\cite{Liang:2023jfj}, yields  $|\bar{\theta}| \lesssim 1.2\times10^{-10}$.
A more recent lattice calculation using a new technique
 gives $d_n = -(5.0\pm0.9)\times10^{-16} \,\bar{\theta}\, e\unit{cm}$~\cite{Blum:2026hul}, 
which corresponds to $|\bar{\theta}| \lesssim 3.6\times10^{-11}$ using the central value of the lattice result.
Although further refinement of lattice QCD calculations for the $\bar\theta$-induced nEDM is still needed,
these results clearly indicate that $|\bar{\theta}|\lsim 10^{-10}$. This extreme smallness of the strong $CP$ violation remains unexplained within the SM and constitutes the strong $CP$ problem.
Several solutions have been proposed, including the Peccei-Quinn mechanism with the QCD axion~\cite{Peccei:1977hh,Weinberg:1977ma,Wilczek:1977pj}, as well as  models based on $CP$~\cite{Nelson:1983zb,Barr:1984qx,Barr:1984fh} or parity symmetry \cite{Beg:1978mt,Mohapatra:1978fy,Babu:1988mw,Barr:1991qx}.

Among the latter, the left-right symmetric model with the extended gauge group $SU(3)_C \times SU(2)_L \times SU(2)_R \times U(1)_{B-L}$ provides a well-motivated framework 
for realizing the parity solution of the strong $CP$ problem.
A generalized parity symmetry forbids the bare QCD $\theta$ term and constrains the quark mass matrices such that $\bar{\theta}$ vanishes at tree level.
In the minimal realization considered in this paper,
the SM quark masses are generated through the universal seesaw mechanism~\cite{Davidson:1987mh,Davidson:1987tr,Davidson:1989bx,Ranfone:1990jf} involving vector-like $SU(2)$ singlet quarks. 
In this setup, the observed quark-mass hierarchy can be partly obtained from an inverted hierarchy in  the vector-like quark masses,
so that the hierarchy among the fundamental Yukawa couplings can be substantially milder than that among the effective SM Yukawa couplings.
Thus, at tree level, this framework simultaneously provides a parity-based solution to the strong $CP$ problem and a possible origin of the hierarchical structure of the SM quark Yukawa couplings.

Once the parity is spontaneously broken, however, radiative corrections can generate a nonzero $\bar{\theta}$. 
The resulting nEDM constraint places bounds on the masses and flavor structure of the UV model, making an accurate calculation of the radiatively induced $\bar{\theta}$ essential for assessing its viability. 
The loop order at which $\bar{\theta}$ is generated, as well as its magnitude, depends sensitively on the mechanism of the parity breaking.
Previous studies based on the radiatively corrected quark-mass determinant showed that the one-loop quark-mass contribution vanishes in the minimal left-right symmetric model with the universal seesaw and estimated the leading contribution at two loops~\cite{Babu:1989rb,Craig:2020bnv}.
Related estimates were also made in high-scale Higgs-parity realizations~\cite{Hall:2018let}.
A two-loop analysis of radiative corrections to $\bar{\theta}$ studied both general \textit{non-Hermitian} vector-like-quark mass matrices and the Hermitian limit relevant to the present work~\cite{deVries:2021pzl}.
In the non-Hermitian case, 
it found potentially non-decoupling threshold corrections that generically produce a large $\bar{\theta}$.\footnote{Non-Hermitian vector-like-quark mass matrices also induce the quark EDM operators directly at one loop, even though their one-loop correction to $\bar{\theta}$ vanishes. 
These direct EDM contributions decouple as the vector-like-quark masses increase; see Ref.~\cite{Craig:2020bnv} for details.}
On the other hand, in the Hermitian limit the corrections are much smaller and are controlled by the soft parity breaking in the scalar sector.
A direct loop-diagrammatic formulation, which consistently includes heavy-quark threshold effects, later confirmed the cancellation through two-loop order in the direct calculation of the radiative correction to $\bar{\theta}$ and estimated the leading nonzero contribution at three-loop order~\cite{Hisano:2023izx}.
This three-loop counting corresponds to that in two-loop corrections to the quark-mass matrix in the mass-determinant approach.

In this paper, we identify the $CP$-odd rephasing invariants, to which the contributions to $\bar\theta$ are proportional, in the minimal left-right model, assuming the Hermitian vector-like quark mass matrices. 
The non-vanishing and non-decoupling corrections arise from differences in the vacuum expectation values and masses of the $SU(2)_L$ and $SU(2)_R$ Higgs doublets, associated with the soft breaking of parity by dimension-two operators.
We identify that the largest $CP$-odd rephasing invariant appears only in the radiative correction to the phases of the determinants of the SM quark mass matrices at the two-loop level.
We find that the bound on $\bar{\theta}$ requires either nearly degenerate vector-like quark masses, which suppress radiative corrections to $\bar{\theta}$ through a GIM-like mechanism, or strongly suppressed mixing angles introduced in the model.
Such a nearly degenerate spectrum does not support the inverted vector-like-quark mass hierarchy introduced to account for the SM quark-mass hierarchy.
Although Ref.~\cite{deVries:2021pzl} studied the case of the Hermitian vector-like quark mass matrices, 
it assumed  degenerate vector-like quark masses except for the top-quark partner. 
We find that the largest contribution to $\bar\theta$ may arise from \textit{nondegenerate vector-like down-type quark masses}, leading to more stringent constraints from the bound on $\bar\theta$.

This paper is organized as follows. In Sec.~\ref{sec:LRmodel}, we briefly introduce the minimal left-right symmetric model and the parametrization of its Yukawa couplings.
It is shown that six phases and six mixing angles are introduced in the model in addition to the CKM phase and mixing angles.
In Sec.~\ref{sec:method}, we formulate the evaluation of $\bar{\theta}$ from the phases of the quark mass matrices and organize the $CP$-odd rephasing invariants that can contribute to the radiative corrections to $\bar{\theta}$, and specify the contributions evaluated in this work.
In Sec.~\ref{sec:result}, we present the numerical results for $\bar{\theta}$ and show that the parameter region realizing an inverted hierarchy of vector-like-quark masses while satisfying the nEDM constraint is severely restricted. Section~\ref{sec:conclusions} is devoted to conclusions and discussion.

\section{The minimal left-right symmetric model}
\label{sec:LRmodel}

\begin{table}[t]
  \centering
  \renewcommand{\arraystretch}{1.3}
  \rowcolors{2}{gray!15}{white}
  \begin{tabular}{ c c c c c | c }
    \bhline{1 pt}

    &
    $SU(3)_C$ &
    $SU(2)_L$ &
    $SU(2)_R$ &
    $U(1)_{B-L}$ &
    $U(1)_Y$

    \\ \hline
    
    $Q_L^i = (u_L^i,d_L^i)^T$ &
    $\mathbf{3}$ &
    $\mathbf{2}$ &
    $\mathbf{1}$ &
    $1/6$ & 
    $(1/6,1/6)$

    \\

    $Q_R^i = (u_R^i,d_R^i)^T$ &
    $\mathbf{3}$ &
    $\mathbf{1}$ &
    $\mathbf{2}$ &
    $1/6$ & 
    $(2/3,-1/3)$

    \\

    $U_L^a$ &
    $\mathbf{3}$ &
    $\mathbf{1}$ &
    $\mathbf{1}$ &
    $2/3$ &
    $2/3$

    \\

    $U_R^a$ &
    $\mathbf{3}$ &
    $\mathbf{1}$ &
    $\mathbf{1}$ &
    $2/3$ &
    $2/3$

    \\

    $D_L^a$ &
    $\mathbf{3}$ &
    $\mathbf{1}$ &
    $\mathbf{1}$ &
    $-1/3$ &
    $-1/3$

    \\

    $D_R^a$ &
    $\mathbf{3}$ &
    $\mathbf{1}$ &
    $\mathbf{1}$ &
    $-1/3$ &
    $-1/3$

    \\

    $H$ &
    $\mathbf{1}$ &
    $\mathbf{2}$ &
    $\mathbf{1}$ &
    $1/2$ &
    $(1/2,1/2)$

    \\

    $H'$ &
    $\mathbf{1}$ &
    $\mathbf{1}$ &
    $\mathbf{2}$ &
    $1/2$ &
    $(1,0)$

    \\

    \bhline{1 pt}
  \end{tabular}
  \caption{Matter content and gauge charges in the minimal left-right symmetric model with the universal seesaw. The indices $i$ and $a$ denote the doublet and singlet quark flavors, respectively, and the $U(1)_Y$ charge is given by $Y=T^3_R+(B-L)$.}
  \label{tab:contents}
\end{table}

We begin with a brief review of the minimal left-right symmetric model.
To restore parity, the Standard Model gauge group is extended to $SU(3)_C \times SU(2)_L \times SU(2)_R \times U(1)_{B-L}$, with $SU(2)_R$ gauge bosons $W'$ and an $SU(2)_R$-doublet Higgs field $H'$.
In addition, the $SU(2)_L$-singlet right-handed quarks $u_R$ and $d_R$ are embedded in the $SU(2)_R$ doublet $Q_R=(u_R,d_R)^T$, and vector-like quarks $U_{L,R}$ and $D_{L,R}$ are introduced to generate the SM quark masses through the universal seesaw mechanism. 
Table~\ref{tab:contents} summarizes the quark and Higgs sectors and their gauge charges.
Furthermore, with this field content, the parity transformation is extended as follows:
\begin{align}
  \vec{x} &\longleftrightarrow -\vec{x}\,, \notag \\
  W^\mu &\longleftrightarrow W'_\mu\,, \\
  Q_L,\ U_L,\ D_L, H &\longleftrightarrow Q_R,\ U_R,\ D_R, H'\,. \notag
\end{align}
The difference between the vacuum expectation values (VEVs), $v^\prime \neq v$ with $\langle H^{(\prime)}\rangle=v^{(\prime)}$, 
spontaneously breaks the extended parity symmetry as well as
breaks $SU(2)_R \times U(1)_{B-L}$ down to $U(1)_Y$, where the hypercharge is given as $Y=T_R^3+(B-L)$.
Subsequently, the electroweak-scale VEV 
breaks $SU(2)_L \times U(1)_Y$ to  $U(1)_{\rm em}$.
We now discuss the Yukawa interactions, the quark mass terms and the Higgs sector in detail.
The phenomenology of the left-right model with a universal seesaw has been studied through flavor observables \cite{Ranfone:1990jf,Kiyo:1998zm,Dcruz:2023mvf}, electric dipole moments \cite{Babu:1989rb}, vacuum stability \cite{Mohapatra:2014qva} and  collider and gravitational-wave searches  \cite{Craig:2020bnv}.

First, we explain the Yukawa interactions and the quark masses.
The Yukawa interactions and Dirac mass terms of the vector-like quarks are given by 
\begin{align}
  -\mathcal{L}_Y
  &= \bar{Q}_L^ix_u^{ia}U_R^a\tilde{H} + \bar{Q}_R^ix_u^{ia}U_L^a\tilde{H}' + M_u^a\bar{U}_L^aU_R^a \notag\\
  &\quad  + \bar{Q}_L^ix_d^{ia}D_R^aH + \bar{Q}_R^ix_d^{ia}D_L^aH' + M_d^a\bar{D}_L^aD_R^a + \text{h.c.}\,,
\end{align}
and the up- and down-type quark mass matrices are
\begin{align}
  \bar{\mathcal{M}}_q =
  \begin{pmatrix}
    0 & vx_q\\
    v^\prime x_q^\dagger& M_q
  \end{pmatrix},
  \quad \text{for $q=u,d$,}
\end{align}
where $i,a=1,2,3$ are flavor indices of doublet and vector-like quarks, respectively, and $\tilde{H}^{(\prime)}=\varepsilon H^{(\prime)*}$.
To respect the parity symmetry, $x_{u,d}^{ia}$ must be the same complex matrix for interactions with both $H$ and $H'$ (the first two terms in each row), and each vector-like mass matrix $M_q$, whose real eigenvalues are given by $M_q^a$  $(q=u,d)$, must be Hermitian.
Furthermore, one can always choose $v$ and $v'$ to be real and positive 
by $SU(2)_L$ and $SU(2)_R$ transformations.
Consequently,
$\det\bar{\mathcal{M}}_q$ is real in each quark sector.
Together with $\theta_G =0$ from the parity symmetry, this yields $\bar{\theta}=0$ at tree level even when the parity is softly broken.
In addition, we assume $v^\prime \lesssim M_q^a$ to explain the light SM quark mass $m_q \ll v$ and its hierarchy via the universal seesaw mechanism.
The SM quark mass is given by
\begin{align}
    m_q = x_q vv^\prime M_q^{-1} x_q^\dagger\,,
    \label{eq:mqformula}
\end{align}
and therefore, the hierarchy among the $x_q$ is expected to roughly follow that among the square roots of the corresponding quark masses. An inverted vector-like quark hierarchy, $M_q^1>M_q^2>M_q^3$, also allows the SM quark hierarchy to be reproduced with a milder hierarchy among the entries of $x_q$. 

Next, we describe a parametrization of the Yukawa couplings in which the quark masses, mixing angles, and $CP$-violating phase are fixed to their observed values.
We work in the weak basis, where the $SU(2)_L$ and $SU(2)_R$ gauge interactions are diagonal.
For convenience, we choose the following basis for $x_q$ and $M_q$ \cite{Hisano:2023izx}:
\begin{equation}
\begin{aligned}
  x_u
  &= V^\dagger_\text{CKM}\sqrt{\frac{m_u}{v}}V_U^\prime\sqrt{\frac{M_u}{v^\prime}}\,, \\
  x_d
  &= \sqrt{\frac{m_d}{v}}V_D^\prime\sqrt{\frac{M_d}{v^\prime}}\,,
\end{aligned}
\label{eq:yukawa}
\end{equation}
where $m_q= m_{qi}\delta_{ij}$,  
$M_q= M_{q}^a\delta_{ab}$, and $V_\text{CKM}$ is the CKM matrix.\footnote{
Equation~\eqref{eq:mqformula} is an approximate formula for the case of $v^\prime \lesssim M_q^a$. Thus,  when $v^\prime\simeq M_q^a $,  the Yukawa couplings in Eq.~\eqref{eq:yukawa}, especially  $(x_u)_{33}$, are modified.
However, we use Eq.~\eqref{eq:yukawa} even when $M_q^a\simeq v^\prime$ in this paper since the formula is much simpler. We consider that these modifications do not change our results qualitatively.
}
The unitary matrices $V_{U/D}^\prime$ are given by $V^\prime$ parametrized as
\begin{align}
  &\ V^\prime(\theta_{12},\theta_{23},\theta_{13},\delta,\theta_3,\theta_8) \notag\\
  &= \exp\left( i\tau_3\theta_3 \right)\exp\left( i\tau_8\theta_8 \right)V(\theta_{12},\theta_{23},\theta_{13},\delta) \notag\\
  &=
  \begin{pmatrix}
      e^{i(\theta_3+\frac{\theta_8}{\sqrt{3}})} & 0 & 0 \\ 
      0 & e^{i(-\theta_3+\frac{\theta_8}{\sqrt{3}})} & 0 \\ 
      0 & 0 & e^{-i\frac{2\theta_8}{\sqrt{3}}}
  \end{pmatrix}
  \begin{pmatrix} 
      c_{12}c_{13} & s_{12}c_{13} & s_{13}e^{-i\delta} \\ -s_{12}c_{23}-c_{12}s_{23}s_{13}e^{i\delta} & c_{12}c_{23}-s_{12}s_{23}s_{13}e^{i\delta} & s_{23}c_{13} \\ s_{12}s_{23}-c_{12}c_{23}s_{13}e^{i\delta} & -c_{12}s_{23}-s_{12}c_{23}s_{13}e^{i\delta} & c_{23}c_{13} \end{pmatrix}\,,
\label{eq:mixingmatrix}
\end{align}
where $c_{ij}=\cos\theta_{ij}$ and $s_{ij}=\sin\theta_{ij}$ $(i,j=1,2,3)$.
Here, $\tau_3$ and $\tau_8$ are the third and eighth Gell-Mann matrices, while $V(\theta_{12},\theta_{23},\theta_{13},\delta)$ is a CKM-like unitary matrix parametrized by three mixing angles and one $CP$-violating phase. Thus each of $V_U^\prime$ and $V_D^\prime$ contains three angles and three physical $CP$-phases in the chosen basis. 
This choice of basis can always be made by redefining the fields through a unitary transformation. Notice that the $CP$-violating phases in $V_{U/D}^\prime$ become unphysical when  $M_q^a$ (a=1,2,3) are degenerate. Thus, the 
radiative correction to $\bar{\theta}$ may receive the GIM-like suppression in a limit of degenerate vector-like quark masses.

Finally, we mention the Higgs sector in this model.
The Higgs potential is given by
\begin{align}
  V(H, H^\prime) = -M_H^2\left( |H|^2 + |H^\prime|^2 \right) + \lambda\left( |H|^2 + |H^\prime|^2 \right)^2 + \kappa\left( |H|^4 + |H^\prime|^4 \right) + \mu^2|H|^2\,.
\end{align}
The first three terms are parity symmetric, whereas $\mu^2$ softly breaks parity.
When $M_H^2>0$, $H$ and $H^\prime$ acquire non-trivial VEVs.
For $\mu^2=0$, either a parity-preserving vacuum, $v=v^\prime$, or a spontaneously parity-breaking vacuum, $v=0,v^\prime\neq0$ or $v\neq0,v^\prime=0$, is realized.
In order to obtain phenomenologically viable parity breaking vacuum $v^\prime \gg v > 0$, it is required that $\mu^2>0$ and $\lambda,\kappa>0$.
Note that the vacuum $v^\prime \gg v > 0$ can be realized without the soft-breaking term through the fine-tuning of the quartic couplings and the top-sector radiative correction~\cite{Hall:2018let}. 
Furthermore, the left-right symmetry must be gauged in the Lagrangian 
 in order to solve the strong $CP$ problem \cite{Benabou:2025viy}.
 We assume that it is realized in the UV theory, and we do not discuss the origin of the soft-breaking term further.

\section{Radiative corrections to the $\theta$ parameter}
\label{sec:method}

First, we review radiative corrections to the $\bar\theta$ parameter arising from  higher-loop diagrams. 
In Refs.~\cite{Hisano:2023izx,Banno:2024apv,Banno:2025pfq}, the authors of these papers developed a method to evaluate such  radiative corrections in the diagrammatic approach by using the Fock-Schwinger gauge. 
It was shown that a one-loop diagram involving a colored fermion,
whose loop momentum is comparable to the fermion mass, 
contributes to $\bar{\theta}$.
This contribution is 
given by $-\theta_q$, where $\theta_q$ denotes the phase of the colored-fermion mass, in accordance with Eq.~(\ref{eq1}) \cite{Hisano:2023izx}. 
In Ref.~\cite{Banno:2024apv}, 
we adopted a toy model in which a $CP$-violating Yukawa coupling between colored fermions and a scalar boson is introduced, and evaluated the radiative corrections to $\bar\theta$ within the model. 
We constructed the effective theory after integrating out the scalar assuming it is heavier than the fermion, and 
compared two evaluations of the radiative corrections to $\bar\theta$ in the effective and original theories. They are almost consistent with each other, including the radiative correction to the fermion mass, and the difference may be identified with a threshold correction from integrating out the heavier particles.

In the minimal left-right symmetric model, there are three contributions to $\bar{\theta}$.
First, radiative corrections to the light (SM) quark mass matrices contribute to $\bar{\theta}$.
The nontrivial contribution to the mass matrices first appears at the two-loop level in the model. 
Second, the quark  chromo-EDM operators are generated by integrating out the vector-like quarks, 
and they induce $\bar{\theta}$ via the renormalization group equation~\cite{Jenkins:2017dyc}. 
The quark chromo-EDM is generated by integration of the vector-like quarks at the two-loop level.  
In this paper, we neglect them because they are suppressed by the heavy vector-like quark masses.
Third, there is the threshold correction to $\bar\theta$ when the vector-like quarks are integrated out.  
We do not evaluate it since a reliable evaluation requires a three-loop calculation. 
We found that it may be subdominant, 
because the $CP$-odd rephasing invariants contributing to $\bar\theta$ via the quark mass corrections 
are larger than those in the threshold corrections as will be shown.

We therefore focus on the two-loop radiative corrections to the quark mass matrices.
After spontaneous parity breaking, the $\bar{\theta}$ parameter receives contributions from phases in the radiatively corrected quark masses,
\begin{align}
  \bar{\theta} 
  &= - \sum_{q=u,d}\text{arg}\, \text{det}\,\mathcal{M}_q 
   = -    \sum_{q=u,d} \Im\Tr\left[ \log \left( 1 +  \bar{\mathcal{M}}_q^{-1}{\delta \mathcal{M}_q}\right)\right]
   \notag\\
  &\simeq -\sum_{q=u,d}\Im\Tr\left[\bar{\mathcal{M}}_q^{-1}{\delta \mathcal{M}_q}\right]\,.
  \label{eq:radiativetheta}
\end{align}
Here, $\mathcal{M}_q$ denotes the full quark mass matrix in either the up-type or down-type quark sector,
while $\delta \mathcal{M}_q$ represents the radiative corrections to the tree-level mass matrix  $\bar{\mathcal{M}}_q$, such that $\mathcal{M}_q=\bar{\mathcal{M}}_q+\delta \mathcal{M}_q$.
At this point, however, there are two issues in evaluating $\bar{\theta}$.
First, in Eq.~\eqref{eq:radiativetheta}, the contributions from SM quarks and vector-like quarks are mixed through the off-diagonal component of $\delta \mathcal{M}$.
Second, although we need $x_q^4$ or higher to construct the $CP$-odd invariant using the Yukawa coupling,
$\bar{\mathcal{M}}_q^{-1}$ contains $x_q^{-1}$.
To clarify these issues, in Sec.~\ref{subsec:generalfomura}, we separate the contributions from the light and heavy quarks, and organize the powers of the Yukawa couplings.
After that, we explicitly identify which Feynman diagrams we need to calculate in order to evaluate the radiative correction to $\bar\theta$.

\subsection{General formula for the radiative corrections via quark mass matrices}
\label{subsec:generalfomura}

Now we discuss the radiative corrections to $\bar\theta$ via quark mass matrices given in Eq.~\eqref{eq:radiativetheta}
in detail. The inverse matrices are
\begin{align}
  \bar{\mathcal{M}}_q^{-1}
  &=
  \begin{pmatrix}
    -\frac{1}{vv^\prime}(x_q^{\dagger})^{-1}_{ic}M^c_q(x_{q})^{-1}_{cj}      
      &\frac1{v^\prime}(x_q^{\dagger})^{-1}_{ib}\\[0.2cm]
      \frac1v(x_{q})^{-1}_{aj} &0
  \end{pmatrix}\,,
\end{align}
and thus the radiative correction ${\delta \mathcal{M}_q}$ contributes to $\bar\theta$ as
\begin{align}
  \bar{\theta} 
  &= -\sum_{q=u,d}{\rm Im~Tr}
   \biggl[-\frac{1}{vv^\prime} \left( (x_q^\dagger)^{-1}M_q x_q^{-1} \right)_{ij}(\delta  \mathcal{M}_q)^{ij} \nonumber \\
   & \qquad +\frac{1}{v^\prime} (x_q^\dagger)^{-1}_{ib}(\delta  \mathcal{M}_q)^{bi}+\frac{1}{v} (x_q)^{-1}_{aj}(\delta  \mathcal{M}_q)^{ja}\biggr]\,. \label{eq:simpleeq}
\end{align}
A subtlety in evaluating $\bar\theta$ concerns the momentum dependence of the off-diagonal two-point functions  $(\delta\mathcal{M}_q)^{bi}$ and $(\delta\mathcal{M}_q)^{ja}$ \cite{Hisano:2023izx}.
When evaluating the radiative correction to $\bar\theta$ in the diagrammatic approach, these two-point functions appear as subdiagrams, and their external quark momentum is integrated over as a loop momentum of the full diagram for $\bar\theta$. The relevant momentum regions differ between the light- and heavy-quark contributions. 
For the light-quark contributions, the external momentum can be neglected relative to the heavy scales entering the two-point functions, whereas this approximation may not be justified for the heavy-quark contributions. To make this distinction explicit, we organize the calculation in terms of the approximate quark mass eigenstates.

Assuming $v\ll v^\prime\lesssim M_q^a$ to explain the SM quark masses via the universal seesaw mechanism, the approximate diagonalization matrices are
\begin{align}
  U_{qL}&\simeq 
  \begin{pmatrix}
    {\boldsymbol I}_3 -\frac12x_q\frac{v^{2}}{M_q^2}x^\dagger_q 
    &
    x_q\frac{v}{M_q}-x_q\frac{v}{M_q} x_q^\dagger x_q \frac{v^{\prime 2}}{M_q^2}-\frac12 x_q\frac{v^2}{M^2_q} x_q^\dagger x_q \frac{v}{M_q}\\[0.2cm]
    -\frac{v}{M_q}x_q^{\dagger}+\frac{v^{\prime 2}}{M_q^2}x_q^\dagger x_q\frac{v}{M_q}x_q^\dagger +\frac12 \frac{v}{M_q}x_q^\dagger x_q\frac{v^2}{M_q^2}x_q^\dagger&{\boldsymbol I}_3-\frac12 \frac{v}{M_q}x^\dagger_q x_q\frac{v}{M_q}
  \end{pmatrix}\,,\\
  U_{qR}&\simeq 
    \begin{pmatrix}
    {\boldsymbol I}_3 -\frac12x_q\frac{v^{\prime 2}}{M_q^2}x^\dagger_q 
    &
    x_q\frac{v^\prime}{M_q}-x_q\frac{v^\prime}{M_q} x_q^\dagger x_q \frac{v^{2}}{M_q^2}-\frac12x_q\frac{v^{\prime 2}}{M_q^2} x_q^\dagger x_q \frac{v^{\prime}}{M_q}\\[0.2cm]
    -\frac{v^\prime}{M_q}x_q^{\dagger}+\frac{v^{2}}{M_q^2}x_q^\dagger x_q \frac{v^\prime}{M_q}x_q^\dagger+\frac12 \frac{v^{\prime}}{M_q}x_q^\dagger x_q \frac{v^{\prime 2}}{M_q^2}x_q^\dagger &{\boldsymbol I}_3-\frac12 \frac{v^\prime}{M_q}x^\dagger_q x_q\frac{v^\prime}{M_q}
  \end{pmatrix}\,, 
\end{align}
so that
\begin{align}
  U^\dagger_{qL}\bar{\mathcal{M}}_q U_{qR}
  &\simeq
  \begin{pmatrix}
    m_{L_q} & \\
    & m_{H_q}
  \end{pmatrix}\,,
\end{align}
with
\begin{align}
  m_{L_q}
  &= -x_q\frac{vv^\prime}{M_q}x^\dagger_q +\frac12 x_q\frac{v^2}{M_q^2}x_q^\dagger x_q \frac{vv^\prime}{M_q} x_q^\dagger+\frac12 x_q\frac{vv^\prime}{M_q}x_q^\dagger x_q \frac{v^{\prime 2}} {M_q^2}x_q^\dagger\,, \\
  m_{H_q}
  &= M_q+\frac12 \frac{v^{2}}{M_q} x_q^\dagger x_q+\frac12 x^\dagger_qx_q\frac{v^{\prime 2}}{M_q}\,.
\end{align}
We next evaluate the correction to the mass eigenvalues induced by $\delta\mathcal{M}_q$.
For the light-quark eigenstates, we parametrize $\delta \mathcal{M}_q$ as
\begin{align}
  \delta \mathcal{M}^L_q
  =
  \begin{pmatrix}
    vv^\prime a_q & vb_q \\
    v^\prime\bar{b}_q & d_q
  \end{pmatrix}\,,
  \label{eq:lightmassmatrix}
\end{align}
which is evaluated from two-point functions with external momentums zero.
On the other hand,
for the heavy-quark eigenstates, 
\begin{align}
  \delta \mathcal{M}^H_q
  =
  \begin{pmatrix}
    vv^\prime \hat a_q & v \hat b_q \\
    v^\prime\hat{\bar{b}}_q & \hat d_q
  \end{pmatrix}\,,
\end{align}
which are evaluated from two-point functions with external momentums around the heavy quark masses.\footnote{More exactly speaking, this correction to $\bar\theta$ should be considered as the threshold correction from three-loop diagrams by integrating out the heavy particles.}

Applying $U_{qL}$ and $U_{qR}$,
the corrections to the light and heavy quark mass matrices, $(\Delta \mathcal{M}_q)^{ij}$ and $(\Delta \mathcal{M}_q)^{ab}$, are given as
\begin{align}
  U^\dagger_{qL}\delta {\mathcal{M}}^L_qU_{qR}
  &\simeq
  \begin{pmatrix}
    (\Delta \mathcal{M}_q)^{ij} & * \\
    * & *
  \end{pmatrix} \,,\nonumber\\
  U^\dagger_{qL}\delta {\mathcal{M}}^H_qU_{qR}
  &\simeq
  \begin{pmatrix}
    * & * \\
    * & (\Delta \mathcal{M}_q)^{ab}
  \end{pmatrix} \,.
\end{align}
In this mass basis, the contribution from the quark masses to $\bar\theta$ is
\begin{align}
  \bar{\theta}
  &= -{\rm Im~Tr}\left[(m_{L_q}^{-1})_{ij} (\Delta \mathcal{M}_q)^{ji}\right]
  -{\rm Im~Tr}\left[(m_{H_q}^{-1})_{ab} (\Delta \mathcal{M}_q)^{ba}\right]\,,
  \label{eq:theta1}
\end{align}
where
\begin{align}
  \Tr\left[(m_{L_q}^{-1})_{ij} (\Delta \mathcal{M}_q)^{ji}\right]
  &=
  -\Tr
  \left[(x_q^\dagger)^{-1}M_q (x_q)^{-1}a_q\right]
  +\Tr
  \left[(x_q)^{-1}b_q-\frac{v^2}{M_q^2}x_q^\dagger b_q\right]
  \nonumber\\
  &
  +\Tr
  \left[(x^\dagger_q)^{-1}\bar{b}_q-x_q \frac{v^{\prime 2}}{M_q^2}\bar{b}_q\right]
  \nonumber\\
  &
  -\Tr
  \left[\frac{1}{M_q}d_q-\frac{v^2}{M_q^2}x_q^\dagger x_q \frac{1}{M_q}d_q-\frac{1}{M_q}x_q^\dagger x_q \frac{v^{\prime 2}}{M_q^2}d_q
  \right]\,, \label{eq:lightpart}\\
  \Tr\left[(m_{H_q}^{-1})_{ab} (\Delta \mathcal{M}_q)^{ba}\right]
  &=
  \Tr\left[\frac{v^2}{M_q^2}x^\dagger_q \hat{b}_q+x_q\frac{v^{\prime 2}}{M_q^2}\hat{\bar{b}}_q\right]
  \nonumber\\
  &
  +\Tr\left[
  \frac{1}{M_q}\hat d_q
  -\frac{v^2}{M^2_q}x^\dagger x_q\frac{1}{M_q}
  \hat d_q-\frac{1}{M_q}x^\dagger x_q\frac{v^{\prime 2}}{M_q^2}
  \hat d_q
  \right]\,. \label{eq:heavypart}
\end{align}
If  $b_q=\hat b_q$, $\bar{b}_q=\hat{\bar{b}}_q$, and $d_q=\hat{d}_q$, 
Eq.~\eqref{eq:theta1} becomes
\begin{align}
 & -\Im\Tr[(m_{L_q}^{-1})_{ij} (\Delta \mathcal{M}_q)^{ji}]
    -{\rm Im~Tr}[(m_{H_q}^{-1})_{ab} (\Delta \mathcal{M}_q)^{ba}]
  \nonumber\\
  &\quad =
  \Im\Tr
  [(x_q^\dagger)^{-1}M_q (x_q)^{-1}a_q]
  -\Im\Tr
  [(x_q)^{-1}b_q]-\Im\Tr
  [(x_q^\dagger)^{-1}\bar{b}_q]\,.
  \label{eq:theta2}
\end{align}
This result, including only the light-quark contribution, is consistent with the weak basis expression 
$-\Tr(\bar{\mathcal{M}}_q^{-1}\delta\mathcal{M}_q)$ in Eq.~\eqref{eq:simpleeq} upon substituting Eq.~\eqref{eq:lightmassmatrix} for $\delta M_q$. The canceled terms would survive when taking into account their dependence on momenta in the light and heavy quark loops. However, it is found that the dominant contributions to $\bar\theta$ have $CP$-odd invariants including $(x_q)^{-1}$, which appear only in the light-quark mass contribution. Therefore, in this paper, we evaluate $\bar{\theta}$ using Eq.~\eqref{eq:theta2}. 

In the above discussion, we assumed $v^\prime\lsim M_q^a$ 
to allow an expansion in $v^\prime/M_q^a$. However, the light-quark contribution remains dominant even when $v^\prime\simeq M_q^a$, since this dominance originates from $CP$-odd invariants involving $(x_q)^{-1}$.

\begin{figure}[p]
    \centering

    \newcommand{\diagfig}[1]{%
        \includegraphics[
            width=0.29\textwidth,
            height=3cm,
            keepaspectratio
        ]{#1}%
    }

    \setlength{\tabcolsep}{2mm}
    \renewcommand{\arraystretch}{1.15}

    \begin{tabular}{@{}ccc@{}}
        \diagfig{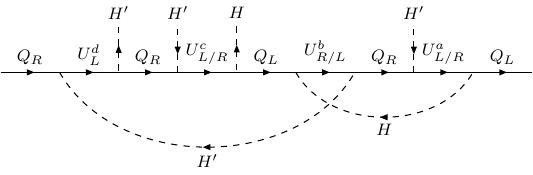} &
        \diagfig{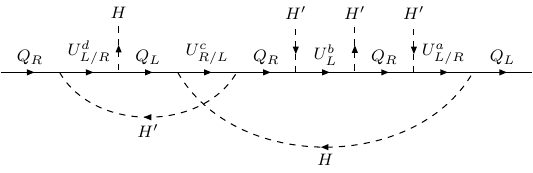} &
        \diagfig{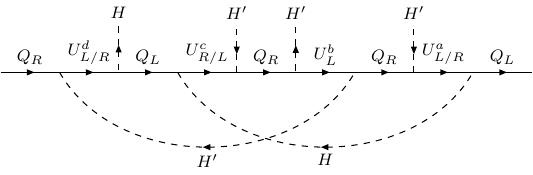}
        \\[3mm]

        \diagfig{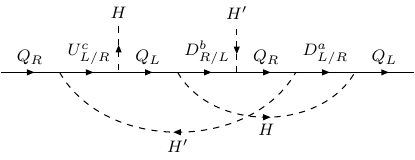} &
        \diagfig{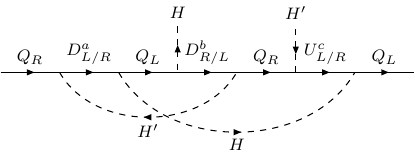} &
        \diagfig{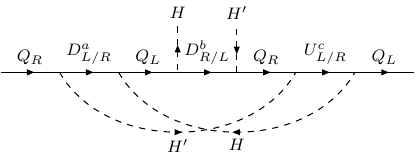}
        \\[3mm]

        \multicolumn{3}{c}{
            \diagfig{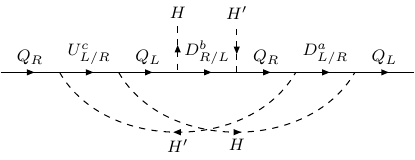}
        }
        \\[3mm]

        \multicolumn{3}{c}{
            \diagfig{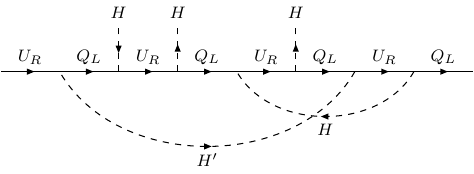}
            \hspace{8mm}
            \diagfig{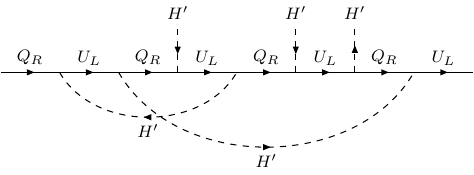}
        }
        \\[5mm]

        \diagfig{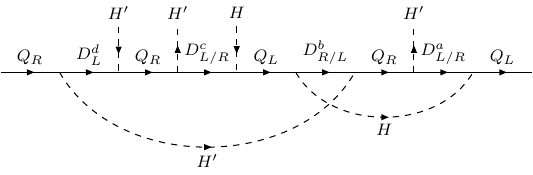} &
        \diagfig{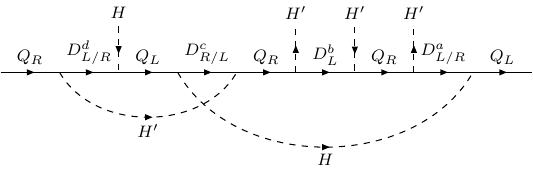} &
        \diagfig{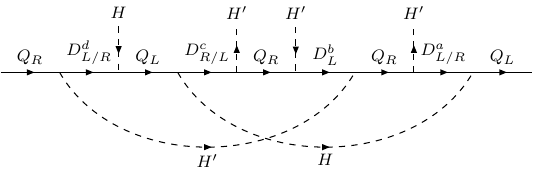}
        \\[3mm]

        \diagfig{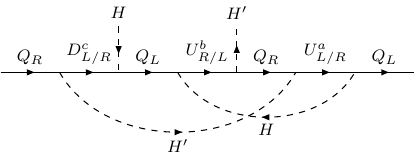} &
        \diagfig{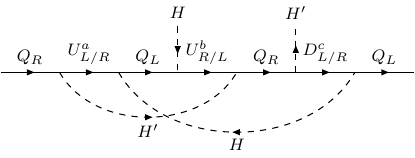} &
        \diagfig{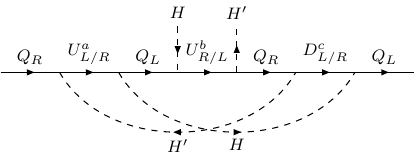}
        \\[3mm]

        \multicolumn{3}{c}{
            \diagfig{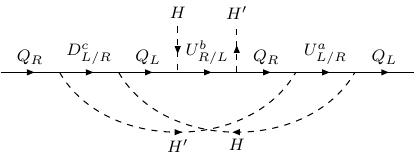}
        }
        \\[3mm]

        \multicolumn{3}{c}{
            \diagfig{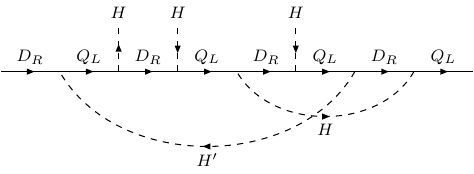}
            \hspace{8mm}
            \diagfig{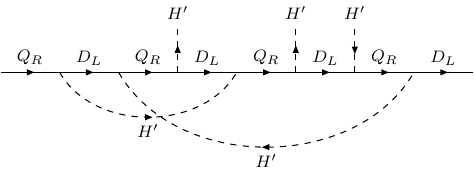}
        }
    \end{tabular}

    \caption{Feynman diagrams corresponding to the contributions to $a_u$ (first to third rows), $b_u$ (fourth row, left), $\bar{b}_u$ (fourth row, right), $a_b$ (fifth to seventh rows), $b_d$ (eighth row, left), and $\bar{b}_d$ (eighth row, right).}
    \label{fig:diagram}
\end{figure}

Figure~\ref{fig:diagram} shows the Feynman diagrams contributing to $a_u$, $b_u$, $\bar{b}_u$ and $a_d$, $b_d$, $\bar{b}_d$. In this paper, we consider the diagrams which consist of the Yukawa couplings, not the gauge coupling and the Higgs quartic coupling. We will discuss the effects of the gauge coupling and the Higgs quartic coupling later. The seven diagrams in the first to third rows correspond to $a_u$; the top three diagrams violate parity via the insertion of $v^\prime$, while the diagrams in the second and third rows violate parity via the mass difference between $H$ and $H^\prime$.
The two diagrams in the fourth row  correspond to $b_u$ and $\bar{b}_u$, and they violate parity via the difference between $v$ and $v^\prime$.
The contributions to $a_d$, $b_d$, and $\bar{b}_d$ can be obtained by interchanging $U$ and $D$ in the diagrams for $a_u$, $b_u$, and $\bar{b}_u$, respectively. They are shown in the fifth to eighth rows in Fig.~\ref{fig:diagram}.
In the following, we neglect $b_u$ and $b_d$, because they are smaller than $\bar{b}_u$ by a factor of $v^2/v^{\prime 2}$.

Next, we consider how the $CP$-odd rephasing invariants, to which the contributions to $\bar\theta$ are proportional, are constructed.
From Fig.~\ref{fig:diagram}, we can see that $a_u \propto (x_ux_u^\dagger)^3, (x_dx_d^\dagger)^2x_ux_u^\dagger$, $\bar{b}_u \propto x_u^\dagger(x_ux_u^\dagger)^3$, $a_d \propto (x_dx_d^\dagger)^3, (x_ux_u^\dagger)^2x_dx_d^\dagger$ and $\bar{b}_d \propto x_d^\dagger(x_dx_d^\dagger)^3$.
Using Eq.~\eqref{eq:theta2}, we obtain the following invariants:
\begin{align}
  &\Im\Tr( [A_u^1,A_u^2] A_u^3 ) \notag\\
  &\quad = -\frac{2J_U}{v^3}(m_t-m_c)(m_t-m_u)(m_c-m_u)\frac{M_u^1M_u^2M_u^3}{v^{\prime 3}}\,, \label{eq:jarlskoguuu} \\
  &\Im\Tr([A_d^1,A_d^2]A_d^3) \notag\\
  &\quad =-\frac{2J_D}{v^3}(m_b-m_s)(m_b-m_d)(m_s-m_d)\frac{M_d^1M_d^2M_d^3}{v^{\prime 3}}\,, \label{eq:jarlskogddd} \\
  &\Im\Tr( A_u^aA_u^bB_d^c ) \notag\\
  &\quad \simeq-\frac{m_t^2}{v^2}\frac{M_u^aM_u^b}{v^{\prime 2}}|V_U^{\prime\,3a}|^2|V_U^{\prime\,3b}|^2
  \bigg[
  \sqrt{\frac{m_b}{m_d}}\Im(V_{td}V_{tb}^*V_D^{\prime\,1c}V_D^{\prime\,3c*})\notag\\
   &\qquad +\sqrt{\frac{m_b}{m_s}}\Im(V_{ts}V_{tb}^*V_D^{\prime\,2c}V_D^{\prime\,3c*})
  \bigg]\,, \label{eq:jarlskoguud}\\
  &\Im\Tr(A_d^aA_d^bB_u^c) \notag\\
  &\quad \simeq-\frac{m_b^2}{v^2}\frac{M_d^aM_d^b}{v^{\prime 2}}|V_D^{\prime\,3a}|^2|V_D^{\prime\,3b}|^2
  \bigg[
  \sqrt{\frac{m_t}{m_u}}\Im(V_{ub}^*V_{tb}V_U^{\prime\,1c}V_U^{\prime\,3c*})\notag\\
  & \qquad  +\sqrt{\frac{m_t}{m_c}}\Im(V_{cb}^*V_{tb}V_U^{\prime\,2c}V_U^{\prime\,3c*})
 \bigg]\,.  \label{eq:jarlskogddu}
\end{align}
The Hermitian matrix $A_q^a$ and the complex matrix $B_q^a$ are defined by
\begin{align}
  (A_q^a)^{ij} &= x_q^{ia}(x_q^\dagger)^{aj}, \\[3pt]
  (B_q^a)^{ij} &= x_q^{ia}(x_q^{-1})^{aj},
\end{align}
where $a$ is not summed. The quantities in  Eqs.~\eqref 
{eq:jarlskoguuu} to \eqref{eq:jarlskogddu} are manifestly invariant under field rephasings. 
$J_U$ and $J_D$ denote the Jarlskog invariants \cite{Jarlskog:1985ht} of $V_U^\prime$ and $V_D^\prime$, defined by
\begin{align}
  \Im\left( V_U^{\prime ia}V_U^{\prime jb}V_U^{\prime *ib}V_U^{\prime *ja} \right) &= J_U\sum_{k,c=1-3}\varepsilon^{ijk}\varepsilon^{abc}, \\
  \Im\left( V_D^{\prime ia}V_D^{\prime jb}V_D^{\prime *ib}V_D^{\prime *ja} \right) &= J_D\sum_{k,c=1-3}\varepsilon^{ijk}\varepsilon^{abc}.
\end{align}

Among the $CP$-phase, $J_U$ and $J_D$ depend on only the phases $\delta$ in Eq.~\eqref{eq:mixingmatrix}
for $V^\prime_U$ and $V^\prime_D$, respectively. Their absolute values are bounded from above by $|J_{U/D}|\leq 1/(6\sqrt{3})\simeq 0.096$, and their mean absolute values are $4/(3\pi^4)\simeq 0.014$ when the mixing angles and $CP$ phases are independently and uniformly sampled over $[0,2\pi]$.
On the other hand, Eqs.~\eqref{eq:jarlskoguud} and \eqref{eq:jarlskogddu} depend on all phases in $V^\prime_D$ and $V^\prime_U$, respectively. They may be nonzero due to the CKM phase even if all phases in $V^\prime_D$ and $V^\prime_U$ are zero. While they are suppressed by the off-diagonal components of the CKM matrix, the suppressions are compensated by square roots of the ratio between heavy and light-quark masses.\footnote{Note that the approximation in Eqs.~\eqref{eq:jarlskoguud} and \eqref{eq:jarlskogddu} may  not be accurate depending on the magnitudes of the mixing angles in $V_U^\prime,V_D^\prime$.}
Comparing the order-of-magnitude estimates by focusing only on 
SM quark masses and CKM matrix elements, we find that $\Im\Tr([A_d^1,A_d^2]A_d^3)$ which appears in $\bar{b}_d$ and $a_d$ is clearly negligible.

The $CP$-odd invariants in Eqs.~\eqref{eq:jarlskoguuu}–\eqref{eq:jarlskogddu} are enhanced by large masses of the first- and second-generation vector-like quarks for the inverted vector-like quark mass hierarchy in the universal seesaw in the universal seesaw, particularly when the mixing angles in $V_U^\prime$ and $V_D^\prime$ are sizable. However, as shown below, larger vector-like quark masses also suppress the loop functions entering their contributions to $\bar{\theta}$. Thus, the larger first- and second-generation vector-like quark masses do not imply a large contribution to $\bar\theta$.

If all vector-like-quark masses are degenerate, the mixing angles and $CP$ phases in $V_U^\prime$ and $V_D^\prime$ are unphysical. Thus, the radiative correction to $\bar\theta$ is suppressed due to the GIM-like mechanism in this case.
To obtain a nonzero contribution to $\bar\theta$, the contributions proportional to  Eqs.~\eqref{eq:jarlskoguuu} and~\eqref{eq:jarlskogddd}, which are completely anti-symmetric under interchange of the vector-like quarks flavor indices, require nondegenerate $U^a$ and $D^a$ masses, respectively. 
The contributions proportional to Eqs.~\eqref{eq:jarlskoguud} and~\eqref{eq:jarlskogddu}, which contain $B_q$, require nondegenerate $D^a$ and $U^a$ masses, respectively, while the $U^a$ and $D^a$ masses, respectively, may be either degenerate or nondegenerate.

In summary, to obtain the leading contributions to $\bar{\theta}$, we need to consider the radiative corrections to $a_q, \bar{b}_q$, which must have the $CP$-odd invariants given by Eqs.~\eqref{eq:jarlskoguuu}, \eqref{eq:jarlskoguud}, and \eqref{eq:jarlskogddu}.

Finally, notice that the anti-symmetric $CP$-odd invariants, such as $\Im\Tr([A_u^1,A_u^2]A_u^3)$, appear in the threshold correction of the vector-like quarks to $\bar\theta$. They are the lowest-order $CP$-odd rephasing invariants constructed from positive powers of the Yukawa couplings \cite{Hisano:2023izx}.
In the next section it will be found that contributions to $\bar\theta$ from the anti-symmetric $CP$-odd invariants via the quark mass matrices are subdominant compared with those including $(B_q^a)^{ij}$, which come from the correction to the light-quark mass matrices. Thus, the other contributions proportional to  the anti-symmetric $CP$-odd invariants
may be negligible. This is also valid for terms we ignore in Eq.~\eqref{eq:heavypart}.

\subsection{Two-loop contributions to quark masses}

As discussed earlier, the quark mass corrections contributing to $\bar{\theta}$ arise at the two-loop level, and we evaluate $\bar{\theta}$ using Eq.~\eqref{eq:theta2}.
For simplicity, we focus on the Feynman diagrams that can be written in the symmetric phase.

We consider the Feynman diagrams that give the leading contributions to $a_u$, $\bar b_u$, and $a_d$. Here, we ignore the contributions proportional to 
the factor $\Im\Tr([A_d^1,A_d^2]A_d^3)$, and we do not show the result for $\bar{b}_d$ and some parts of $a_d$.

After spontaneous symmetry breaking, these contributions take the form
\begin{align}
  a_u
  &= \frac{v^{\prime 2}}{(16\pi^2)^2}\left( A_u^aA_u^bA_u^cA_u^d \right)F_{1, u}^{abcd} \notag\\
  &\ + \frac{1}{(16\pi^2)^2}\left[ \left( A_d^aA_d^bA_u^c \right)F_{3,u}^{ab;c}(m_H,m_{H^\prime})
  + \left( A_u^cA_d^bA_d^a \right)F_{3,u}^{ab;c}(m_{H^\prime},m_H) \right], \\
  \bar{b}_u 
  &= \frac{v^{\prime 2}}{(16\pi^2)^2}\left( x_u^\dagger A_u^aA_u^bA_u^c \right)F_{2,u}^{abc}, \\
  a_d
  &= \frac{1}{(16\pi^2)^2}\left[ \left( A_u^aA_u^bA_d^c \right)F_{3,d}^{ab;c}(m_H,m_{H^\prime}) + \left( A_d^cA_u^bA_u^a \right)F_{3,d}^{ab;c}(m_{H^\prime},m_H) \right],
\end{align}
where $F_{1,u}$, $F_{2,u}$, $F_{3,d}$, and $F_{3,u}$ are loop functions depending on the vector-like-quark flavor indices.
$F_{1,u}$ and $F_{2,u}$ are defined as
\begin{align}
  F_{1,u}^{abcd}
  &= (16\pi^2)^2\int\frac{\mathrm{d}^4p\mathrm{d}^4q}{(2\pi)^8}\frac{p\cdot q}{p^2(p^2-(M_u^a)^2)(p^2-m_H^2)q^2(q^2-(M_u^d)^2)(q^2-m_{H^\prime}^2)} \notag\\
  &\ \times\bigg[ \frac{M_u^aM_u^bM_u^c}{((p-q)^2-(M_u^b)^2)(q^2-(M_u^c)^2)} + \frac{M_u^aM_u^cM_u^d}{(p^2-(M_u^b)^2)((p-q)^2-(M_u^c)^2)} \notag\\
  &\ + \frac{M_u^aM_u^cM_u^d}{((p-q)^2-(M_u^b)^2)((p-q)^2-(M_u^c)^2)} \bigg], \\[5pt]
  F_{2,u}^{abc}
  &= -(16\pi^2)^2\int\frac{\mathrm{d}^4p\mathrm{d}^4q}{(2\pi)^8}\frac{p \cdot q}{p^2(p^2-(M_u^a)^2)(p^2-m_{H^\prime}^2)((p-q)^2-(M_u^b)^2)(q^2-(M_u^c)^2)(q^2-m_H^2)},
\end{align}
and $F_{3,d}$ is
\begin{align}
  F_{3,d}^{ab;c}(m_H,m_{H^\prime})
  &= (16\pi^2)^2\int\frac{\mathrm{d}^4p\mathrm{d}^4q}{(2\pi)^8}\frac{M_u^aM_u^bM_d^c}{(p^2-(M_u^a)^2)((p-q)^2-(M_u^b)^2)(q^2-(M_d^c)^2)} \notag\\
  &\ \times\left[ \frac{q\cdot(q-p)}{q^2(p-q)^2} + \frac{1}{(p-q)^2} \right]\frac{1}{(p^2-m_H^2)(q^2-m_{H^\prime}^2)}.
\label{eq:func}
\end{align}
Regarding $F_{3,d}$, the dependence on the doublet-Higgs masses is displayed explicitly to clarify the parity breaking.
In addition, we can get $F_{3,u}$ by interchanging $M_u$ and $M_d$ in $F_{3,d}$.
The analytic results for general two-loop integrals are given in Refs.~\cite{Ford:1992pn,Davydychev:1992mt,Davydychev:1995mq,Caffo:1998du,Espinosa:2000df},
and analytic expressions for $F_{1,u}$, $F_{2,u}$, and $F_{3,d}$ are provided in the ancillary file \texttt{twoloop\_functions.txt}.

Using Eq.~\eqref{eq:theta2}, the leading contribution to $\bar\theta$ is
\begin{align}
  \bar{\theta} 
  &\simeq \Im\Tr[(x_u^\dagger)^{-1}M_u (x_u)^{-1}a_u] - \Im\Tr[(x_u^\dagger)^{-1}\bar{b}_u] + \Im\Tr[(x_d^\dagger)^{-1}M_d (x_d)^{-1}a_d] \notag\\
  &= \frac{v^{\prime 2}}{(16\pi^2)^2}\frac{1}{2}\Im\Tr\left([A^1_u,A^2_u]A^3_u\right)
  \left[ \sum_{a,b,c} \varepsilon^{abc}\left( M_u^aF_{1,u}^{abca} - F_{2,u}^{abc}  \right) \right] \notag\\
  &\ + \frac{1}{(16\pi^2)^2}\sum_{a,b,c}\Im\Tr\left( A_u^aA_u^bB_d^c \right)M_d^c\left( F_{3,d}^{ab;c}(m_H,m_{H^\prime})-F_{3,d}^{ab;c}(m_{H^\prime},m_H) \right)  \notag\\
  &\ + \frac{1}{(16\pi^2)^2}\sum_{a,b,c}\Im\Tr\left( A_d^aA_d^bB_u^c \right)M_u^c\left( F_{3,u}^{ab;c}(m_H,m_{H^\prime})-F_{3,u}^{ab;c}(m_{H^\prime},m_H) \right). \label{eq:theta3}
\end{align}
Here, the contribution is separated into three parts associated with the three $CP$-odd invariants in Eqs.~\eqref{eq:jarlskoguuu},~\eqref{eq:jarlskoguud}, and ~\eqref{eq:jarlskogddu}.

\section{Results}
\label{sec:result}

\begin{table}[tbp]
  \centering
  \begin{tabular}{cc@{\hspace{2em}}cc}
    \hline\hline
    Parameter & Input value
      & Parameter & Input value \\
    \hline
    $m_u$ & $1.05\,\mathrm{MeV}$
      & $m_c$ & $531\,\mathrm{MeV}$ \\

    $m_d$ & $2.28\,\mathrm{MeV}$
      & $m_b$ & $2.41\,\mathrm{GeV}$ \\

    $m_s$ & $45.1\,\mathrm{MeV}$
      & $m_t$ & $143\,\mathrm{GeV}$ \\
    \hline
    $\sin\theta_{12}$ & $0.22517$
      & $\sin\theta_{23}$ & $0.04189$ \\

    $\sin\theta_{13}$ & $0.003763$
      & $\delta_{\mathrm{CKM}}$ & $1.154$ \\
    \hline\hline
  \end{tabular}
  \caption{Input values of the quark masses and CKM parameters. The quark masses are running masses evaluated at $1\,\mathrm{TeV}$. The mixing angles $\theta_{12}$, $\theta_{23}$, and $\theta_{13}$ and the $CP$-violating phase $\delta_{\rm CKM}$ in the CKM matrix are given in the convention of $V$ in Eq.~\eqref{eq:mixingmatrix}.}
  \label{tab:input-parameters}
\end{table}

We now evaluate the QCD $\theta$ parameter using Eq.~\eqref{eq:theta3}.
For simplicity, we introduce three mass scales $m_1$, $m_2$, and $m_3$ for the vector-like quark masses as
\begin{align}
M_u^1& =M_d^1=M_d^2~(\equiv m_1)\,, \nonumber\\
M_u^2& =M_d^3~(\equiv m_2)\,,\nonumber\\
M_u^3& =v^\prime=m_{H^\prime}~(\equiv m_3)\,.
\end{align}
The vector-like quark masses are expected from the SM quark mass spectrum. If $M_u^3$ is much larger than $v^\prime$, the Yukawa coupling constant $(x_u)^{33}$ is much larger than 1 so that the perturbative  expansion is broken. Then, we take $M_u^3=v^\prime$.
We also assume $v^\prime=m_{H^\prime}$ for simplicity. 
Additionally, we calculate the loop functions by setting $m_H=0$ because it is negligibly small compared with the $W^\prime$ boson mass bound from the search for 
$W^\prime\rightarrow tb$ at the LHC experiment \cite{CMS:2023gte}.
The result then depends only on the ratios among $m_1$, $m_2$, and $m_3$, together with the six mixing angles and six $CP$ phases parameterizing $V^\prime_U$ and $V^\prime_D$.
This is because $\bar\theta$ is dimensionless and the contributions are non-decoupling. As shown in Table~\ref{tab:input-parameters}, the SM parameters entering the Yukawa couplings are taken from PDG~\cite{ParticleDataGroup:2026mpi}, while the quark masses are run to 1 TeV by using \texttt{RunDec}~\cite{Herren:2017osy}.

\begin{figure}[t]
    \centering
    \includegraphics[width=0.45\textwidth]{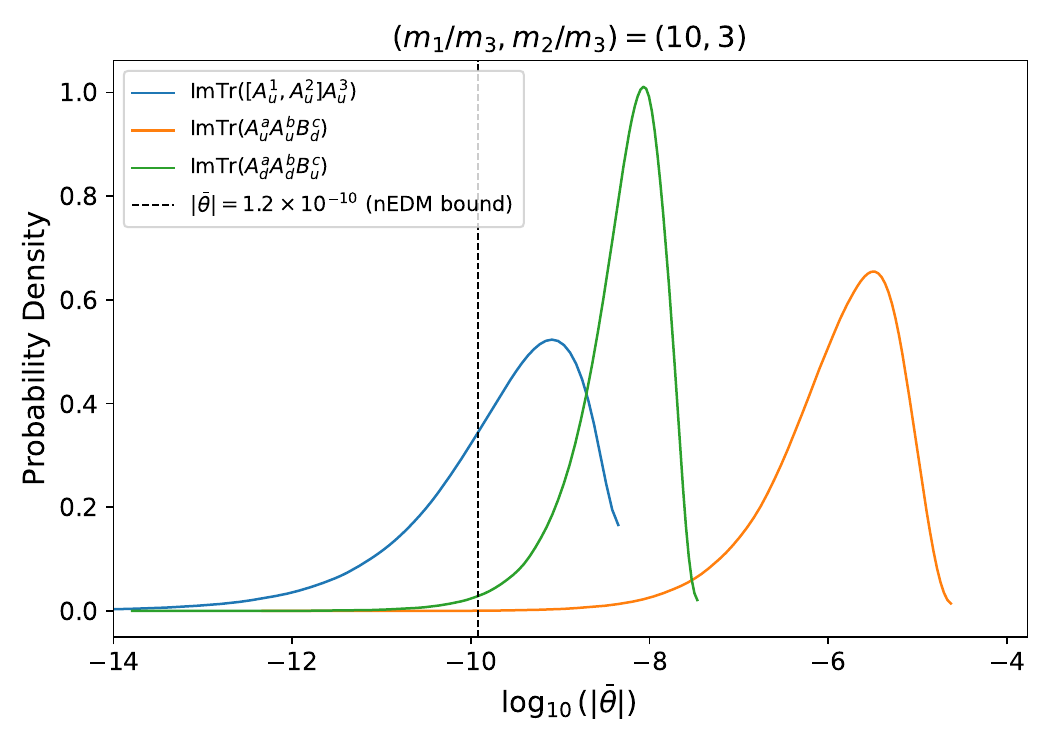} \hfill
    \includegraphics[width=0.45\textwidth]{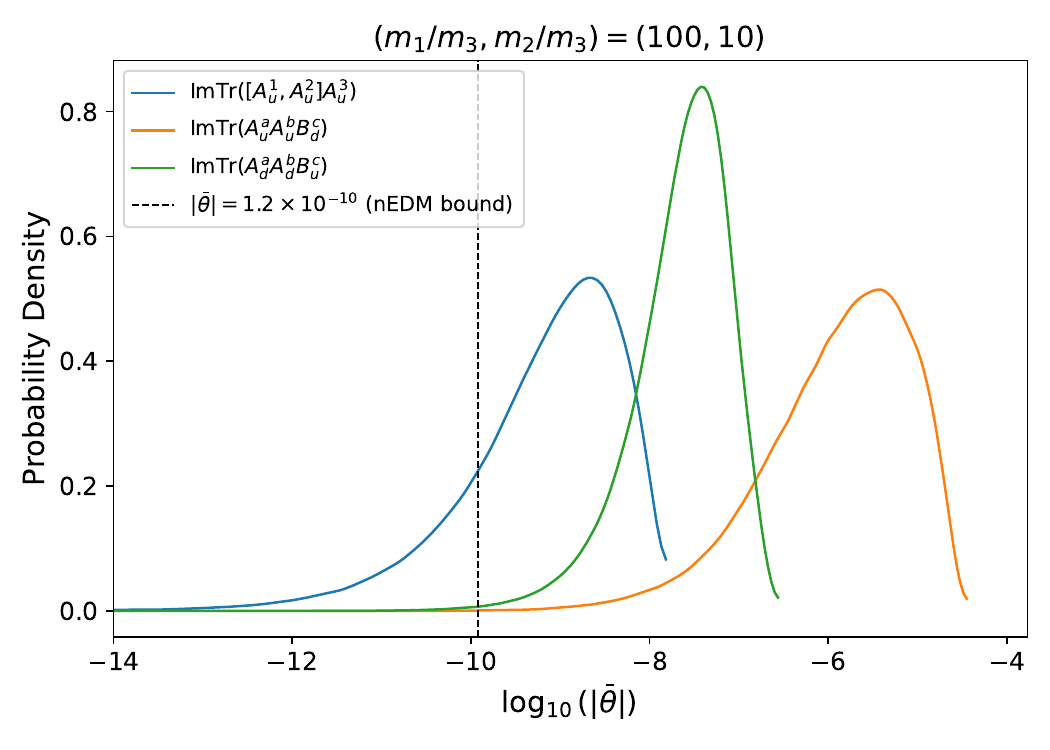}
    \caption{Probability densities of $\log_{10}|\bar\theta|$ for $(m_1/m_3,m_2/m_3)=(10,3)$ (left) and $(100,10)$ (right), restricted to points for which the singular values of $x_u$ and $x_d$ do not exceed $\sqrt{4\pi}$. The blue, orange and green  curves show the three terms in Eq.~\eqref{eq:theta3}. The vertical dashed line corresponds to the reference upper bound from the neutron EDM experiment ($|\bar{\theta}|=1.2\times10^{-10}$). }
    \label{fig:thetaPDF}
\end{figure}

Figure~\ref{fig:thetaPDF} shows the probability density functions (PDFs) of $\log_{10}|\bar\theta|$ for $(m_1/m_3,m_2/m_3)=(10,3)$ and $(100,10)$.
The distributions show the first term in Eq.~\eqref{eq:theta3} in blue, the second term in orange, and the third term in green.
We sample the twelve parameters defining $V^\prime_U$ and $V^\prime_D$, namely six mixing angles and six phases, independently and uniformly over $[0,2\pi]$ and construct the resulting PDFs.
To retain perturbative Yukawa couplings, we exclude parameter points for which the largest singular value of either $x_u$ or $x_d$ exceeds $\sqrt{4\pi}$.
The scan contains one million parameter points for each benchmark.
After imposing the perturbativity constraint on the Yukawa couplings, nearly 100\% and about 20\% of the parameter points remained for $(m_1/m_3,m_2/m_3)=(10,3)$ and $(100,10)$, respectively.

For both benchmark spectra, the total distribution is almost the same as the distribution of $\Im\Tr(A_u^aA_u^bB_d^c)$, and it peaks at $|\bar{\theta}|$ of order $10^{-5} \text{--} 10^{-6}$, above the reference upper bound from the neutron EDM experiments, $|\bar{\theta}|=1.2\times10^{-10}$,  shown in the figure.
Similarly, the peak of the sub-leading distribution, $\Im\Tr(A_d^aA_d^bB_u^c)$, is located at $10^{-7} \text{--} 10^{-8}$ and also exceeds the reference upper bound.
The relative locations of the three component distributions reflect both the typical magnitudes of their $CP$-odd invariants and the mass dependence of the corresponding loop functions.
Their different widths and small-$|\bar{\theta}|$ tails arise from the $CP$-phase dependence of the invariants.
Increasing the hierarchy from $(10,3)$ to $(100,10)$ shifts the peaks of the distributions toward larger $|\bar{\theta}|$ by roughly one order of magnitude.
Thus $\bar{\theta}$ tends to exceed the experimental limits in these benchmarks.

Except for accidental cancellations, there are three possible ways to avoid the upper bound from the neutron EDM by reducing the value of $\bar{\theta}$: 
increasing the masses of all vector-like quarks relative to $v^\prime$ and $m_{H^\prime}$, suppressing the mixing angles in $V_{U}^\prime$ and $V_{D}^\prime$, or approaching a degenerate vector-like-quark spectrum.

\begin{figure}
    \centering
    \includegraphics[width=0.5\linewidth]{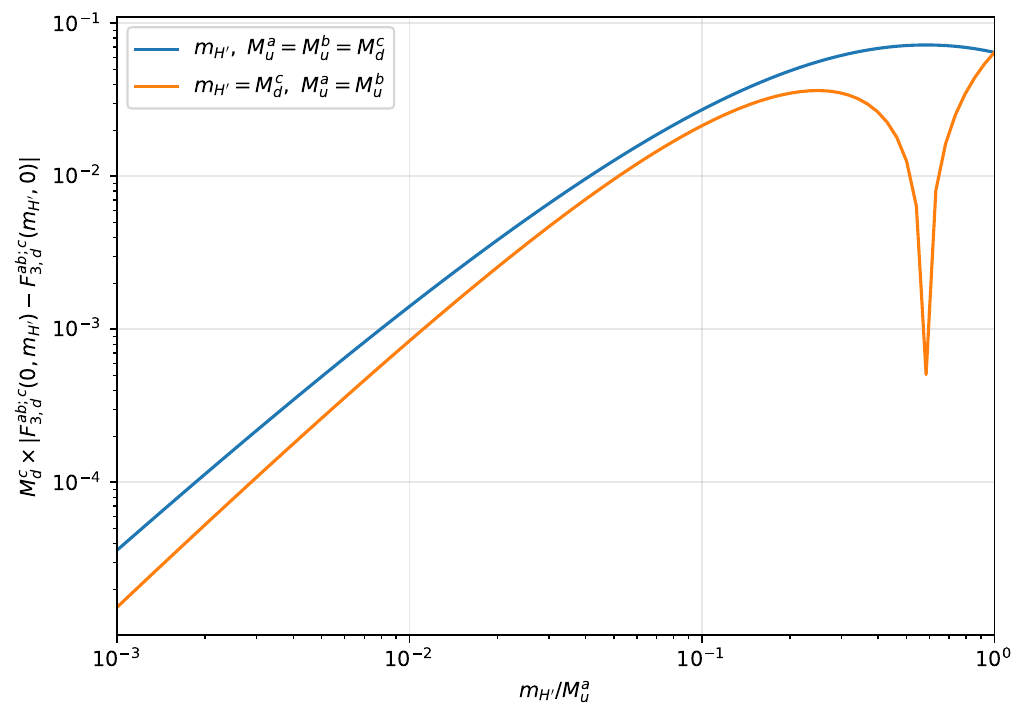}
    \caption{Dependence of $M_d^c\times |(F_{3,d}^{ab;c}(m_H,m_{H^\prime})-F_{3,d}^{ab;c}(m_{H^\prime},m_H))|$ on $m_{H^\prime}/M_u^a$ for $m_H=0$. The blue curve corresponds to $M_u^a=M_u^b=M_d^c$, while the orange corresponds to $m_{H^\prime}=M_d^c$ and $M_u^a=M_u^b$. }
    \label{fig:plotF3}
\end{figure}
The first strategy is strongly limited by perturbativity. $\bar\theta$ is dimensionless so that it depends on the ratios of the dimensionful parameters. It must be suppressed by the parity breaking parameters, $v^\prime$ or $m_{H^\prime}$, if they are smaller than the vector-like quark masses. 
Figure~\ref{fig:plotF3} shows the behavior of the dimensionless loop function $M_d^c\times (F_{3,d}^{ab;c}(m_H,m_{H^\prime})-F_{3,d}^{ab;c}(m_{H^\prime},m_H))$ associated with $\Im\Tr(A_u^aA_u^bB_d^c)$ in Eq.~\eqref{eq:theta3}, which provided the dominant contribution to $\bar{\theta}$ in the above discussion. 
It can be seen that the value of the loop function is suppressed as the vector-like quarks become heavier relative to $H^\prime$.
This behavior is consistent with the fact that the soft parity breaking associated with the mass difference between $m_H\,(v)$ and $m_{H'}\,(v')$ becomes suppressed in this limit.
On the other hand,  Eq.~\eqref{eq:yukawa} shows that increasing the vector-like quark masses enhances the associated Yukawa couplings. Consequently, most sampled points eventually fail the perturbativity bound.

\begin{figure}[t]
    \centering
   \includegraphics[width=0.45\textwidth]{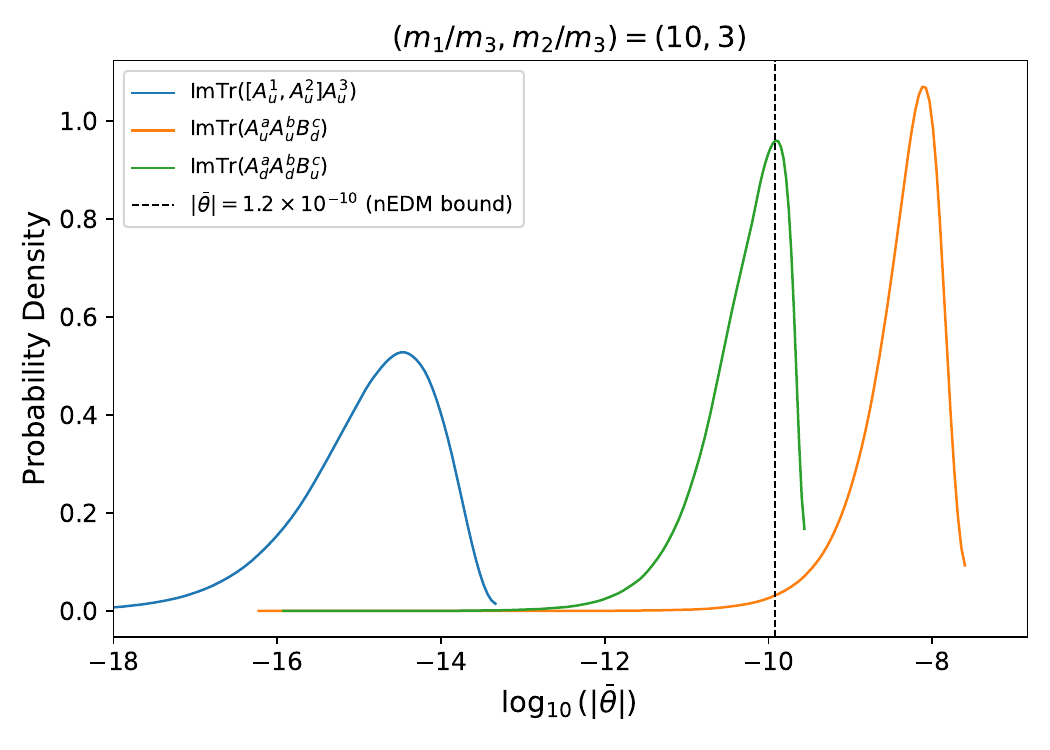} \hfill
    \includegraphics[width=0.45\textwidth]{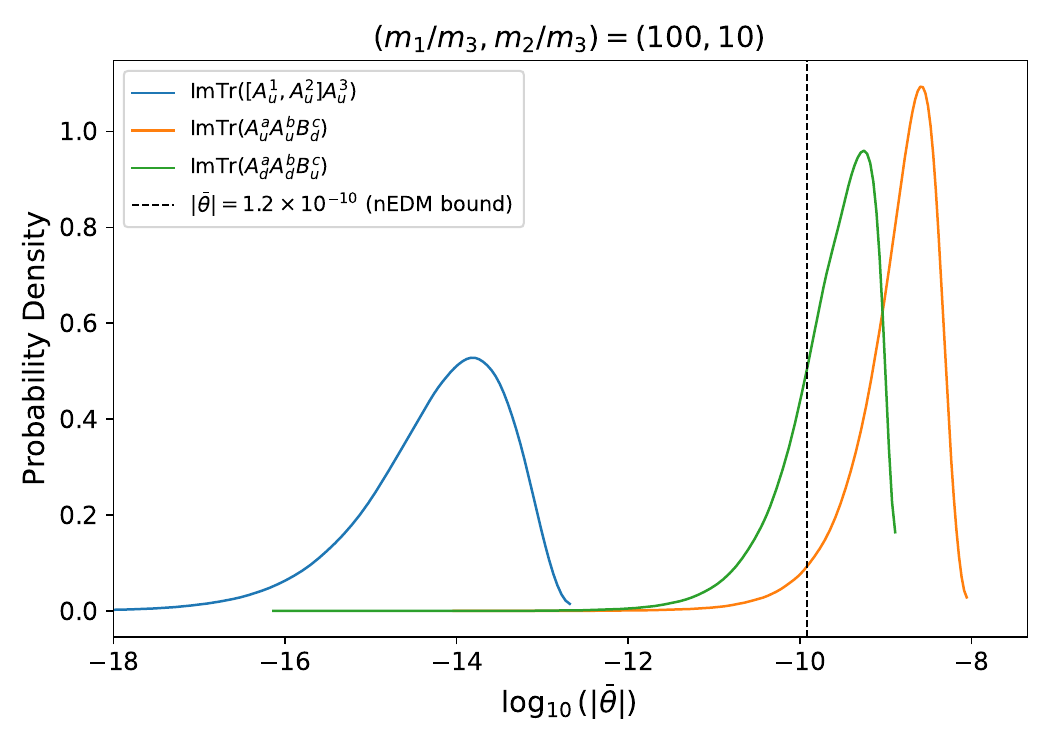}\\
       \includegraphics[width=0.45\textwidth]{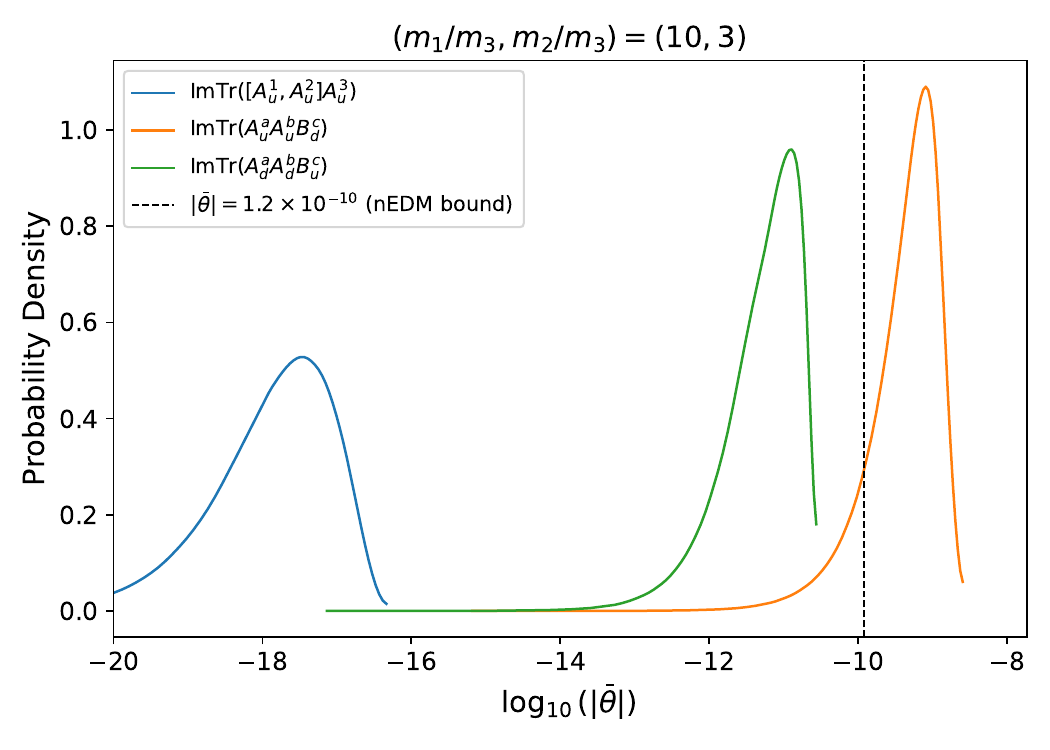} \hfill
    \includegraphics[width=0.45\textwidth]{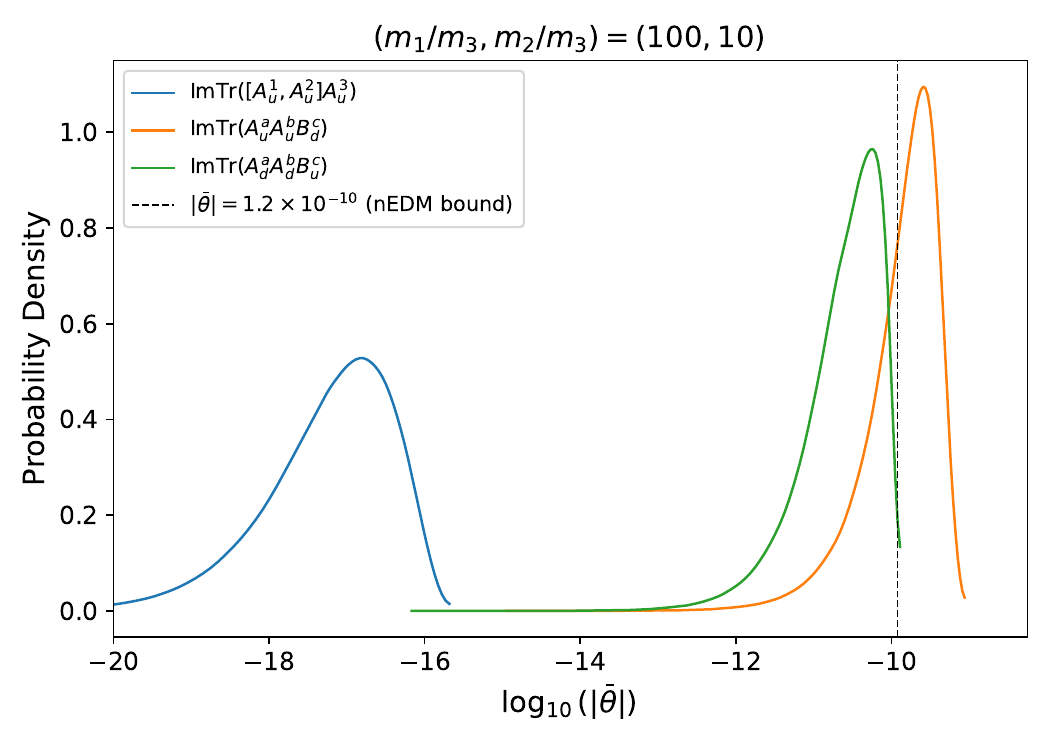}\\
    \caption{The same figures as in Fig.~\ref{fig:thetaPDF}, though the six mixing angles are restricted to the ranges $[-10^{-2},10^{-2}]$ (upper)
    and $[-10^{-3},10^{-3}]$ (lower). }
    \label{fig:thetaPDFc}
\end{figure}
The second strategy is to suppress the mixing angles in $V_{U}^\prime$ and $V_{D}^\prime$.
In Fig.~\ref{fig:thetaPDFc}, PDFs of $\log_{10}|\bar\theta|$ are shown with the same input parameters as Fig.~\ref{fig:thetaPDF}, except that the six mixing angles are restricted to the ranges $[-10^{-2},10^{-2}]$ (upper) and $[-10^{-3},10^{-3}]$ (lower). The contributions of 
$\Im\Tr(A_u^aA_u^bB_d^c)$ and $\Im\Tr(A_d^aA_d^bB_u^c)$
become smaller by a factor of about $10^3$ ($10^4$) in the upper (lower) figures. However, the viable prediction for $\bar\theta$ is limited in those figures. 

Equation~\eqref{eq:jarlskoguud} is proportional to $M_u^aM_u^b |{V_U}^{\prime3a}|^2|{V_U}^{\prime3b}|^2$. The factor may be larger for larger off-diagonal terms in ${V^\prime _U}$ since $M_u^1\ge M_u^2 \ge M_u^3$. However, the larger vector-like quark masses suppress the loop function in Eq.~\eqref{eq:func}, as shown in Fig.~\ref{fig:plotF3}. 
Then, the contributions of 
$\Im\Tr(A_u^aA_u^bB_d^c)$ are suppressed by only a mixing in $\Im(V_{td}V_{tb}^*V_D^{\prime\,1c}V_D^{\prime\,3c*})$ and 
$\Im(V_{ts}V_{tb}^*V_D^{\prime\,2c}V_D^{\prime\,3c*})$. Similarly, the contributions of $\Im\Tr(A_d^aA_d^bB_u^c)$ are also suppressed by only a mixing in $\Im(V_{ub}^*V_{tb}V_U^{\prime\,1c}V_U^{\prime\,3c*})$ and 
$\Im(V_{cb}^*V_{tb} V_U^{\prime\,2c}V_U^{\prime\,3c*})$.
Then, their contributions to $\bar\theta$ are suppressed by only one power of the mixing angle as shown in the figures.

The third option suppresses $\bar{\theta}$ through the GIM-like cancellation by taking degenerate vector-like quark masses. When assuming $(m_1/m_3,m_2/m_3)=(1+2\Delta,1+\Delta)$ and taking $\Delta\ll1$, 
we find that
the \mbox{$\Im\Tr([A^1_u,A^2_u]A^3_u)$} part scales as $\Delta^3$ in Eq.~\eqref{eq:theta3}, while the \mbox{$\Im\Tr( A_u^aA_u^bB_d^c )$} and \mbox{$\Im\Tr( A_d^aA_d^bB_u^c )$} parts scale as $\Delta$.

\begin{figure}[t]
    \centering
    \includegraphics[width=0.45\textwidth]{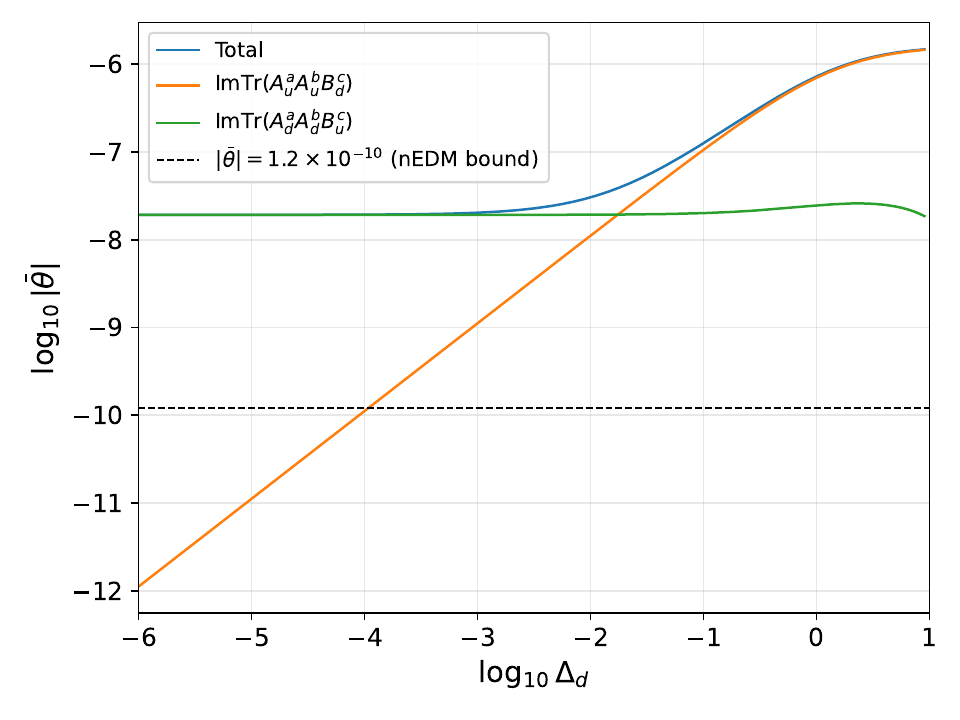}
    \hfill
    \includegraphics[width=0.45\textwidth]{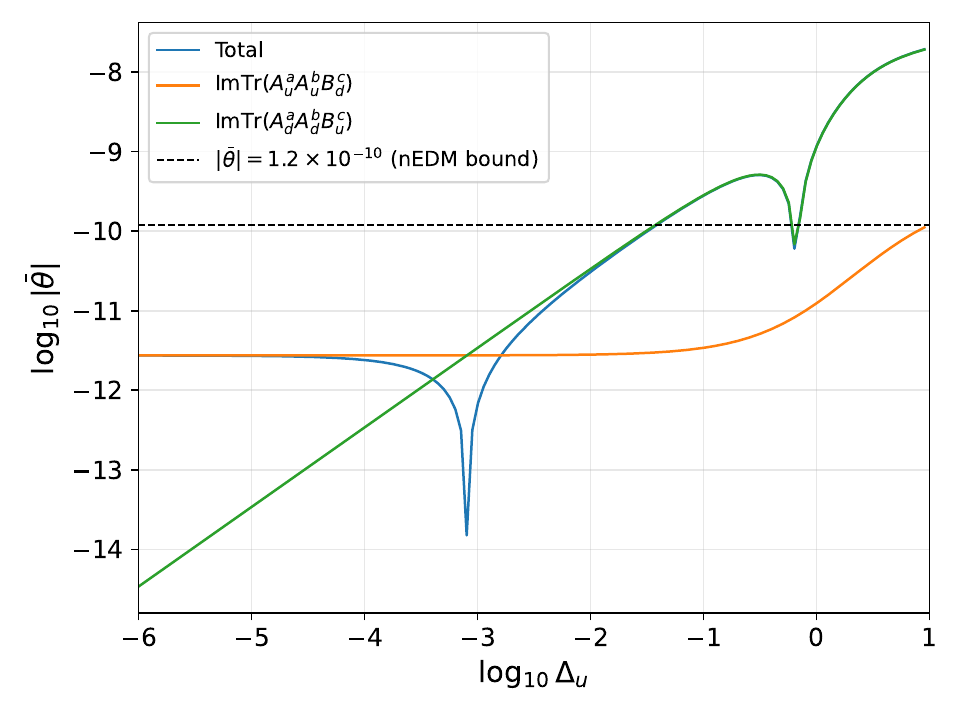}
    \caption{The change in $\bar{\theta}$ when $\Delta_d$ or $\Delta_u$ is decreased. 
    The left (right) panel corresponds to fixed $\Delta_u=9$ ($\Delta_d=10^{-4}$).
    We adopted the angular parameters near the peak shown in the right panel of Fig.~\ref{fig:thetaPDF}: $V_U^\prime=V' (6.08, 5.11, 3.73, 2.41, 1.66, 4.07),\ V_D^\prime=V' (3.72, 4.76, 0.191, 3.60, 4.59, 0.0254)$. The horizontal dashed line corresponds to the reference upper bound from the neutron EDM experiment ($|\bar{\theta}| = 1.2\times 10^{-10}$).}
    \label{fig:degenerate}
\end{figure}

\begin{figure}
    \centering
    \includegraphics[width=0.5\linewidth]{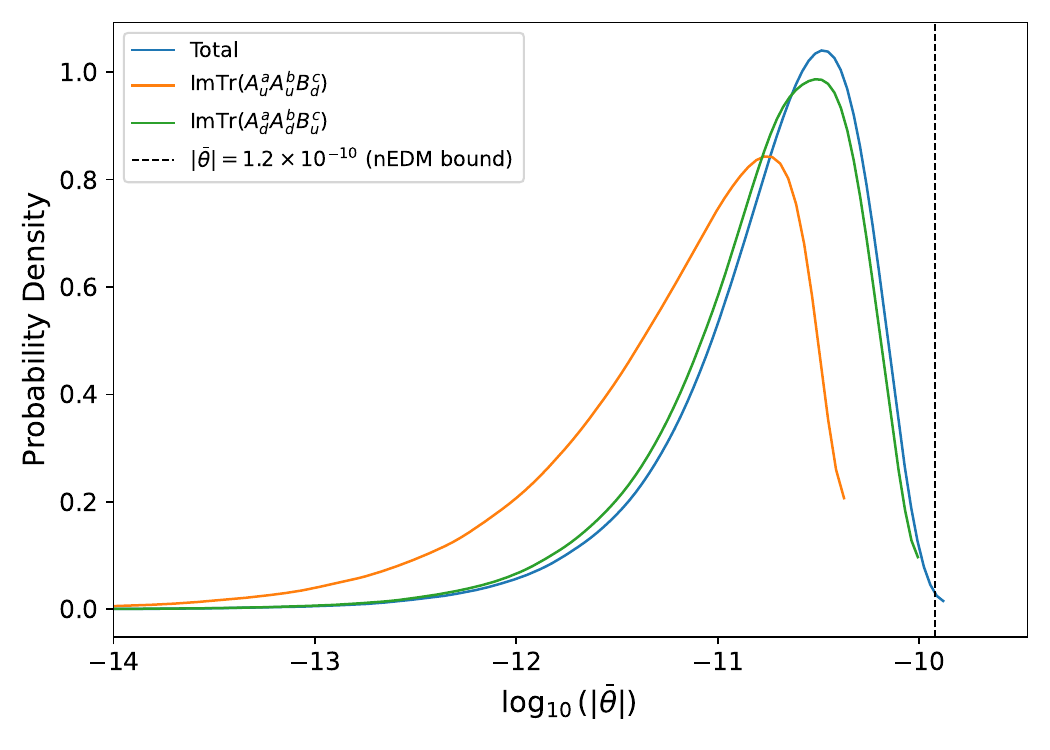}
    \caption{Probability densities of $\log_{10}|\bar\theta|$ for $\Delta_u=10^{-2},\Delta_d=10^{-4}$, restricted to points for which the singular values of $x_u$ and $x_d$ do not exceed $\sqrt{4\pi}$. The blue curve shows the total distribution, while the orange and green curves show the two terms in Eq.~\eqref{eq:theta3}. The vertical dashed line corresponds to the reference upper bound $|\bar{\theta}|=1.2\times 10^{-10}$ from the current experiment.}
    \label{fig:deltaPDF}
\end{figure}

We clarify how degenerate the masses need to be in order to obtain values below the experimental constraints.
We introduce parameters $\Delta_u$ and $\Delta_d$ to represent the degree of degeneracy,
\begin{align}
\begin{aligned}
M_u^1/M_u^2&=M_u^2/M_u^3=1+\Delta_u\,,  \\
M_d^1/M_d^3&=M_d^2/M_d^3=1+\Delta_d\,, 
\end{aligned}
\end{align}
and $v^\prime=M_u^3=M_d^3/10$.
For example, when $\Delta_u=\Delta_d=9$, the result agrees with the second benchmark $(m_1/m_3,m_2/m_3)=(100,10)$.
As can be seen on the right side of Fig.~\ref{fig:thetaPDF}, the contribution from \mbox{$\Im\Tr( A_u^aA_u^bB_d^c )$} significantly exceeds the experimental constraint. 
We therefore first consider reducing it by decreasing $\Delta_d$. 
Here, we fixed the mixing angles and $CP$ phases in $V^\prime_U$ and $V^\prime_D$ as a sample point specified by $V^\prime
_U = V^\prime (6.08,5.11,3.73,2.41,1.66,4.07)$, $V^\prime_D =
V^\prime (3.72,4.76,0.191,3.60,4.59,0.0254)$ ($V'$ is given in Eq.~\eqref{eq:mixingmatrix}). The left panel of Fig.~\ref{fig:degenerate} illustrates the change in $\bar{\theta}$ when $\Delta_d$ is reduced with fixed $\Delta_u=9$.
It shows that $\Delta_d<10^{-4}$ is required for the contribution from \mbox{$\Im\Tr( A_u^aA_u^bB_d^c )$} to remain below the experimental bound.
We next consider the contribution from \mbox{$\Im\Tr( A_d^aA_d^bB_u^c )$}, and the right panel of Fig.~\ref{fig:degenerate} illustrates the change in $\bar{\theta}$ when $\Delta_u$ is reduced with $\Delta_d=10^{-4}$.
It shows that $\Delta_u \lesssim 10^{-2}$ is also required for $\bar{\theta}$ to remain below the experimental bound.
The peak around $\log_{10}\Delta_u\sim0$ comes from accidental cancellation among contributions proportional to $\Im\Tr( A_d^aA_d^bB_u^c )$, while the peak around $\log_{10}\Delta_u\sim-3$ arises from the cancellation between contributions from
$\Im\Tr( A_d^aA_d^bB_u^c )$ and $\Im\Tr( A_u^aA_u^bB_d^c )$.
Figure~\ref{fig:deltaPDF} shows the PDFs of $\log_{10}|\bar\theta|$ for $\Delta_u=10^{-2},\Delta_d=10^{-4}$.
Here, the contribution from $\Im\Tr( [A_u^1,A_u^2]A_u^3 )$ is neglected because it decreases sufficiently rapidly as $\Delta_u$ decreases.

In the above calculation, we used the leading-order approximation in $v'/M_q^a$. Although this approximation may  not be valid for the vector-like top quark, satisfying the nEDM constraint still requires tuning the up-type and down-type vector-like quark mass parameters to precisions better than approximately $1\%$ and $0.01\%$, respectively, even if the higher-order corrections in the expansion of $v'/M_q^a$ are included.

Finally, let us discuss the other contributions to $\bar\theta$ via the SM quark mass matrices in the left-right symmetric model. We evaluated the corrections to the SM quark mass matrices from the Yukawa interactions and then derived the contribution to $\bar\theta$ in the model. Other diagrams may also contribute to the SM quark mass matrices at the two-loop level when the gauge couplings and the Higgs-self couplings are switched on \cite{Hall:2018let, deVries:2021pzl}. Their contributions to $\bar\theta$ 
are proportional to the lower-order invariants, given by
\begin{align}
    \Im\Tr(A_u^aB_d^b)
    &\simeq-\frac{m_t}{v}\frac{M_u^a}{v^\prime}|V_U^{\prime\,3a}|^2
    \bigg[
    \sqrt{\frac{m_b}{m_d}}\Im(V_{td}V_{tb}^*V_D^{\prime\,1b}V_D^{\prime\,3b*}) \notag\\
    &\quad +\sqrt{\frac{m_b}{m_s}}\Im(V_{ts}V_{tb}^*V_D^{\prime\,2b}V_D^{\prime\,3b*})
    \bigg]\,, \label{eq:jarlskogud} \\
    \Im\Tr(A_d^aB_u^b)
    &\simeq-\frac{m_b}{v}\frac{M_d^a}{v^\prime}|V_D^{\prime\,3a}|^2
    \bigg[
    \sqrt{\frac{m_t}{m_u}}\Im(V_{ub}^*V_{tb}V_U^{\prime\,1b}V_U^{\prime\,3b*}) \notag\\
    &\quad +\sqrt{\frac{m_t}{m_c}}\Im(V_{cb}^*V_{tb}V_U^{\prime\,2b}V_U^{\prime\,3b*})
    \bigg]\,. \label{eq:jarlskogdu}
\end{align}

We restrict ourselves to an order-of-magnitude estimate in this paper.
Let us assume that all the vector-like quark masses are nearly degenerate: $M_u^1 \sim M_u^2 \sim M_u^3 \sim M_d^1 \sim M_d^2 \sim M_d^3 \sim v^\prime$ since it is expected from the above discussion.
Since $\bar{\theta}$ is dimensionless and there is only one mass scale under the above assumption, when gauge couplings or the Higgs quartic coupling are involved, $\bar{\theta}$ can be estimated as follows:
\begin{align}
    \bar{\theta}
    &\simeq \frac{1}{(16\pi^2)^2}\left[ \Im\Tr( A_u^aA_u^bB_d^c ) +  g^2\Im\Tr( A_u^aB_d^b ) \right]\Delta_d \notag\\
    &\ + \frac{1}{(16\pi^2)^2}\left[\Im\Tr( A_d^aA_d^bB_u^c ) + g^2\Im\Tr( A_d^aB_u^b ) \right]\Delta_u \notag\\
    &\simeq \frac{1}{(16\pi^2)^2}\left( \frac{m_t^2}{v^2} + g^2\frac{m_t}{v} \right)
    \bigg[
    \sqrt{\frac{m_b}{m_d}}\Im(V_{td}V_{tb}^*V_D^{\prime\,1b}V_D^{\prime\,3b*})  +\sqrt{\frac{m_b}{m_s}}\Im(V_{ts}V_{tb}^*V_D^{\prime\,2b}V_D^{\prime\,3b*})
    \bigg]\Delta_d \notag\\
    &\ + \frac{1}{(16\pi^2)^2}\left( \frac{m_b^2}{v^2} + g^2\frac{m_b}{v} \right)
    \bigg[
    \sqrt{\frac{m_t}{m_u}}\Im(V_{ub}V_{tb}^*V_U^{\prime\,1b}V_U^{\prime\,3b*})  +\sqrt{\frac{m_t}{m_c}}\Im(V_{cb}V_{tb}^*V_U^{\prime\,2b}V_U^{\prime\,3b*})
    \bigg]\Delta_u ,
\end{align}
where $\Delta_u$ and $\Delta_d$ denote the mass splittings of the up- and down-type vector-like quarks, and we removed $|V^{\prime 3a}_{U,D}|$ using $\sum_{a=1}^3|V^{\prime 3a}_{U,D}|=1$.
Additionally, $g^2$ represents any gauge coupling in the model, but it can be regarded as the Higgs quartic coupling.
Assuming that the combination of $V_\mathrm{CKM}$ and $V^\prime_{U,D}$ satisfies $\Im(V_{ij}V_{ik}^*V_{U,D}^{\prime\,jb}V_{U,D}^{\prime\,kb*}) \sim |V_{ij}V_{ik}^*|\times 10^{-1}$, the above expression can be numerically estimated as
\begin{align}
    \bar{\theta} \sim (0.8 + g^2)\times10^{-6}\times\Delta_d + (0.01 + g^2)\times10^{-7}\times\Delta_u \,.
\end{align}
This estimate indicates that when constraining the up-type mass splitting $\Delta_u$, the gauge contributions cannot be neglected. For $g^2 \simeq 0.3$, 
the gauge contribution alone gives $|\bar{\theta}| \sim 3 \times 10^{-8}\times \Delta_u$, which exceeds the nEDM bound for $\Delta_u = 10^{-2}$. 
Thus, including the gauge contributions could strengthen the constraint on $\Delta_u$ obtained from the right panel of Fig.~\ref{fig:degenerate}.
A quantitative determination of this constraint requires an explicit calculation of these contributions.

\section{Conclusions}
\label{sec:conclusions}

The left-right symmetric model is motivated by a solution of the strong $CP$ problem, and also by the universal seesaw's  explanation of the quark mass hierarchy. The latter favors the inverted hierarchy of the vector-like quark masses. We studied the $CP$-odd two-loop radiative corrections to quark masses, which contribute to the QCD $\theta$ parameter, in the minimal left-right symmetric model with a non-universal vector-like quark mass spectrum.
After organizing the contributions of the quark mass matrices to $\bar{\theta}$ and the $CP$-odd invariants, we conducted a two-loop calculation of the diagrams that give the leading contributions.
For benchmark spectra with hierarchical vector-like-quark masses, the resulting probability distributions of $\bar{\theta}$ show that most of the sampled points exceed the current nEDM bound.
Apart from suppressing the mixing angles in the model, $\bar\theta$ can be suppressed via the GIM-like mechanism by making the vector-like quark masses nearly degenerate.
We find that satisfying the nEDM constraint requires tuning the up-type and down-type vector-like quark mass parameters to precisions better than approximately $1\%$ and $0.01\%$, respectively.
Therefore, the parameter region in which the inverted vector-like-quark mass hierarchy motivated by the universal seesaw can be realized while satisfying the nEDM constraint is severely restricted.

\acknowledgments

We are sincerely grateful to Prof.~Yukinari Sumino for providing us with intensive lectures on
the details of higher-order loop calculations. This work is supported by the JSPS Grant-in-Aid for Scientific Research Grant No.22K21347
(T.K.), No.24K07016 (J.H.), No.24K22872
(T.K.), No.25H02180 (J.H.), and No.25K07276 (T.K.).
The work of J.H. is also supported by World Premier
International Research Center Initiative (WPI Initiative), MEXT, Japan. T.B. is supported
by the “Make New Standards Program for the Next Generation Researchers” of the Tokai
National Higher Education and Research System (THERS). This work is also financially
supported by JST SPRING, Grant Number JPMJSP2125.

\appendix

\bibliographystyle{utphys28mod}
\bibliography{ref}

\end{document}